\documentclass{article}

\usepackage{microtype}
\usepackage{graphicx}
\usepackage{subcaption}
\usepackage{booktabs} 

\usepackage{hyperref}

\usepackage{xurl}

\usepackage[preprint]{icml2026}

\usepackage{amsmath}
\usepackage{amssymb}
\usepackage{mathtools}
\usepackage{amsthm}

\usepackage[capitalize,noabbrev]{cleveref}

\crefname{appendix}{Appendix}{Appendices}
\Crefname{appendix}{Appendix}{Appendices}

\usepackage{algorithm}
\usepackage{algorithmic}

\theoremstyle{plain}

\theoremstyle{definition}

\theoremstyle{remark}

\icmltitlerunning{Predictor-Driven Diffusion for Spatiotemporal Generation}

\begin{document}

\twocolumn[
  \icmltitle{Predictor-Driven Diffusion for Spatiotemporal Generation}



  \icmlsetsymbol{equal}{*}

  \begin{icmlauthorlist}
    \icmlauthor{Yuki Yasuda}{jamstec}
    \icmlauthor{Tobias Bischoff}{aeolus}
  \end{icmlauthorlist}

  \icmlaffiliation{jamstec}{Research Institute for Value-Added-Information Generation, Japan Agency for Marine-Earth Science and Technology, Kanagawa, Japan}
  \icmlaffiliation{aeolus}{Aeolus Labs, San Francisco, United States}

  \icmlcorrespondingauthor{Yuki Yasuda}{yuki.yasuda@jamstec.go.jp}

  \icmlkeywords{diffusion models, renormalization group, surrogate modeling, dynamical systems, generative models, path integrals}

  \vskip 0.3in
]



\printAffiliationsAndNotice{}  

\begin{abstract}
    Multiscale spatial structure complicates temporal prediction because small-scale spatial fluctuations influence large-scale evolution, yet resolving all scales is often intractable. Standard diffusion models do not address this problem effectively since they apply uniform decay across all Fourier modes. We propose Predictor-Driven Diffusion, a framework that combines renormalization-group-based spatial coarse-graining with a path-integral formulation of temporal dynamics. The forward process applies scale-dependent Laplacian damping together with additive noise, producing a hierarchy of coarse-grained fields indexed by diffusion scale $\lambda$. Training minimizes the Kullback--Leibler divergence between data-induced and predictor-induced path densities, leading to a simple regression loss on temporal derivatives. The resulting predictor captures how eliminated small-scale components statistically influence large-scale evolution. A key insight is that the same predictor provides a path score for reverse-$\lambda$ sampling, unifying simulation, unconditional generation, and super-resolution in a single framework. Our unified approach is validated through experiments on two multiscale turbulent systems.
\end{abstract}

\section{Introduction}

Many natural and engineered systems exhibit hierarchical structure, where phenomena at different levels of detail influence each other \citep{Levin92,BarYam+04,Newman+12,Alexakis+Biferale18}. For example, weather patterns are shaped by smaller-scale processes such as cloud formation \citep{Bony+15}. In signal processing, Laplacian pyramids exploit this structure by decomposing images into coarse components plus progressively finer residuals \citep{Burt+87,Moulin+09}. Such hierarchical representations are valuable for modeling and prediction because different levels interact \citep{Mallat89,Weinan+11,Bengio+13}.

One way to quantify this hierarchy is through spatial Fourier decomposition \citep{Hussaini+Zang87,Zhang+Lu04}: since wavenumber is inversely proportional to wavelength, modes at small and large wavenumbers capture large-scale and small-scale structure, respectively \citep[e.g.,][]{Bracewell00}. A system is spatially \emph{multiscale} when it contains Fourier modes spanning a wide range of wavenumbers. This Fourier-based view provides a useful foundation for data analysis and machine learning on multiscale data \citep{Rahimi+Recht07,Mallat+16,Li+21}.

Diffusion models have emerged as powerful tools for generation \citep{Ho+20,Song+21}, in part because their coarse-to-fine denoising appears to use multiscale structure \citep{Karras+22,Saharia+22a}. In standard diffusion, small-scale components tend to be corrupted faster in terms of signal-to-noise ratio, so the reverse (generation) process proceeds from large-scale to small-scale structure \citep{Rissanen+23,Falck+25}. However, this multiscale hierarchy arises implicitly from noise: the model denoises all Fourier modes simultaneously rather than separating scales explicitly. This means standard diffusion does not fully exploit spatial multiscale structure \citep{Sheshmani+25,Masuki+Ashida25}.

In physics, the renormalization group (RG) is a traditional method to address multiscale structure \citep{Wilson71,Berges+02}. RG builds a fine-to-coarse hierarchy by progressively integrating out small-scale degrees of freedom while preserving their statistical influence on large-scale components. Recent work has combined RG with diffusion models, incorporating scale-aware damping and noise into the forward process \citep{Cotler+Rezchikov23,Sheshmani+25,Masuki+Ashida25}. The reverse (generation) process then naturally corresponds to super-resolution---reconstructing small-scale structure from coarse representations.

Modeling dynamical systems is a fundamental task in physics and engineering, yet existing RG-based diffusion models have focused on static data such as images \citep{Cotler+Rezchikov23,Sheshmani+25,Masuki+Ashida25}. Extending RG-based diffusion to spatiotemporal dynamics is not straightforward. Coarse-graining (or smoothing) along the physical time axis would mix future information into the present, violating causality \citep[e.g.,][]{Einicke06,Einicke12}. A key open problem is to achieve this extension while preserving both causality and explicit scale awareness.

We address this problem by proposing \emph{Predictor-Driven Diffusion}. The key idea is to learn a predictor of forward evolution in physical time and use it to define a probability distribution over spatiotemporal trajectories across multiple spatial scales. The predictor computes future states only from past observations, so the framework preserves causality by construction. This approach enables diffusion-like sampling across spatial scales, unifying simulation, generation, and super-resolution within a single model.

Our main contributions are:
\begin{itemize}
    \item A joint treatment of physical time (causal evolution) and spatial scale (RG-based hierarchy) as distinct axes, producing coarse-grained fields at multiple resolutions (\cref{sec:background,sec:proposed-method}).
    \item A training objective based on KL divergence between trajectory distributions, which theoretically yields a predictor capturing the influence of small scales on large-scale evolution (\cref{sec:proposed-method}).
    \item A unified framework where a network trained for simulation also supports unconditional generation and super-resolution without retraining (\cref{sec:proposed-method,sec:experiments}).
\end{itemize}
The framework integrates naturally with existing predictor architectures, requiring minimal modifications. We demonstrate these capabilities on chaotic systems with one- and two-dimensional (1D and 2D) spatial domains.

\begin{figure*}[t]
    \vskip 0.2in
    \begin{center}
        \includegraphics[width=\textwidth]{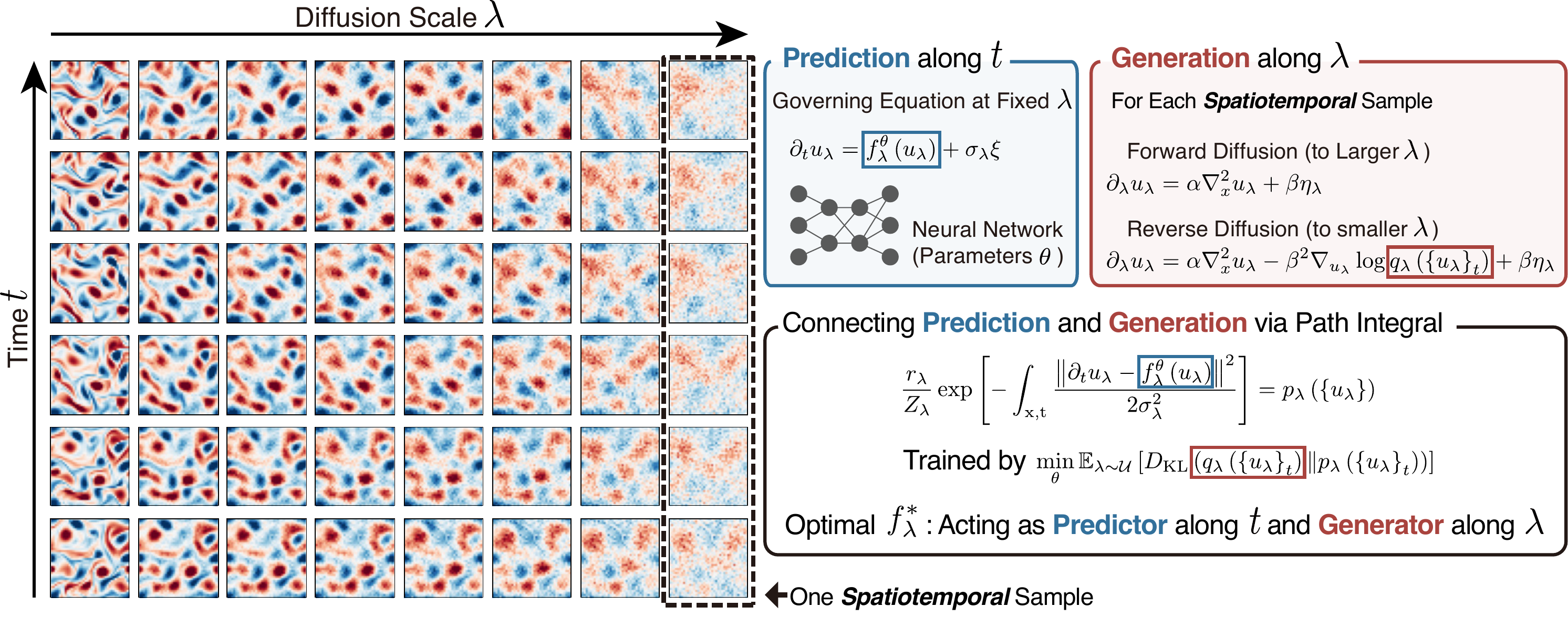}
        \caption{Schematic of the proposed framework. The horizontal axis represents the diffusion scale $\lambda$, which controls the degree of coarse-graining; the vertical axis represents physical time $t$. Blue box: forward \emph{simulation} in $t$ emulates dynamics at any fixed $\lambda$; red box: reverse-$\lambda$ integration of a spatiotemporal sample enables \emph{generation} from noise and \emph{super-resolution} from a coarse-grained input. A single predictor, trained by minimizing KL divergence between path densities (bottom box), supports all three tasks (i.e., simulation, generation, and super-resolution).}
        \label{fig:schematic}
    \end{center}
\end{figure*}

\section{Background}
\label{sec:background}

We review recent work that combines RG coarse-graining with diffusion models \citep[e.g.,][]{Cotler+Rezchikov23}. This review is presented in terms of two axes: physical time $t$, which indexes causal evolution, and diffusion scale $\lambda$ (defined below), which indexes a spatial scale hierarchy. Prior RG-based diffusion has focused on static fields, where only the $\lambda$ axis is present. We restate those results with explicit dependence on $t$ to prepare for the spatiotemporal extension in \cref{sec:proposed-method}.

\subsection{RG Transformations via Diffusion Processes}
\label{subsec:rg-transformation-via-diffusion-processes}

RG coarse-graining can be formulated as a stochastic diffusion process in $\lambda$ \citep{Carosso20}. This is one formulation of RG; we adopt it for concreteness. For a broader perspective on RG approaches, see \citet{Cotler+Rezchikov23}.

For a vector field $u_\lambda := u_\lambda(x,t)$, the forward process is
\begin{equation}
\label{eq:forward-rg-sde}
    \partial_\lambda u_\lambda = \alpha \nabla_x^2 u_\lambda + \beta \eta_\lambda,
\end{equation}
where $\alpha, \beta > 0$ are constants, $\nabla_x^2$ is the spatial Laplacian, and $\eta_\lambda := \eta_\lambda(x,t)$ is white Gaussian noise ($\mathbb{E}[\eta_\lambda(x,t)\eta_{\lambda'}(x',t')] = \delta(\lambda-\lambda')\delta(x-x')\delta(t-t')$). In Fourier space, the mode at wavenumber $k$ is attenuated by a factor $\exp(-\alpha \left\|k\right\|^2 \lambda)$, so modes with large $\left\|k\right\|$ (small-scale structure) decay faster than those with small $\left\|k\right\|$ (large-scale structure); as $\lambda$ increases, $u_\lambda$ becomes progressively coarse-grained. A cutoff wavenumber $k_{\mathrm{cut}} \sim 1/\sqrt{\alpha\lambda}$ separates retained modes from suppressed ones \citep{Carosso20}. Since the cutoff wavelength is proportional to $\sqrt{\lambda}$, the diffusion scale $\lambda$ indexes a spatial hierarchy (\cref{subsec:coarse-graining-operator-and-covariance-matrix}).

The noise term in Eq.~(\ref{eq:forward-rg-sde}) is essential for RG coarse-graining. Without noise ($\beta = 0$), the Laplacian alone performs deterministic smoothing, which simply removes small-scale components \citep{Carosso20}. With noise, the corresponding Fokker--Planck equation shows that the forward process induces a convolution of the field distribution with a Gaussian kernel (\cref{subsec:coarse-graining-operator-and-covariance-matrix,subsec:carosso-renormalization-group}), corresponding to marginalization over small-scale degrees of freedom \citep{Wetterich91,Carosso20}. This marginalization preserves the statistical effect of eliminated components on large-scale components, distinguishing RG coarse-graining from simple low-pass filtering.

\subsection{Learning the Inverse RG Transformation}

Given the forward process in Eq.~(\ref{eq:forward-rg-sde}), the corresponding reverse process in $\lambda$ takes the form \citep{Anderson82,Song+21}
\begin{equation}
\label{eq:reverse-rg-sde}
    \partial_\lambda u_\lambda = \alpha \nabla_x^2 u_\lambda - \beta^2 \nabla_{u_\lambda} \ln q_\lambda(\{u_\lambda\}_t) + \beta \eta_\lambda.
\end{equation}
Here, $q_\lambda(\{u_\lambda\}_t)$ denotes the coarse-grained path density at diffusion scale $\lambda$. We use $\{u_\lambda\}_t$ to denote the spatiotemporal \emph{path} (field values at all grid points) and $\nabla_{u_\lambda}$ the gradient with respect to these field values. This reverse process generates large-scale structure first and progressively restores small-scale components as we integrate from large to small $\lambda$. \citet{Cotler+Rezchikov23} proposed RG-based diffusion: learning the score function via denoising score matching enables generation as \emph{inverse} RG coarse-graining, which is interpreted as super-resolution.

Scale-dependent damping is essential for interpreting the reverse process as super-resolution. Following \citet{Cotler+Rezchikov23}, we adopt Laplacian-based damping, one of several possible scale-dependent forms \citep{Sheshmani+25,Masuki+Ashida25}. Standard diffusion models, such as variance-preserving formulations \citep{Song+21}, apply scale-uniform damping and noise across all Fourier modes; thus, intermediate states are organized by noise level rather than spatial resolution \citep{Kingma+21,Falck+25}. While the reverse process may appear to recover structure from large to small scales, this ordering arises from differences in the signal-to-noise ratio, not from an explicit spatial hierarchy \citep{Falck+25,Masuki+Ashida25}. In contrast, diffusion with scale-dependent damping guarantees generation from large to small scales, exploiting the hierarchical structure of the data \citep{Sheshmani+25,Masuki+Ashida25}.

RG-based diffusion has been studied for static fields \citep{Cotler+Rezchikov23,Sheshmani+25,Masuki+Ashida25}. Extending to dynamical systems requires a probability density over trajectories that respects causal evolution: future states must depend only on past states. To construct such a density, we employ a path-integral formulation based on a learned predictor.

\section{Proposed Method}
\label{sec:proposed-method}

We propose \emph{Predictor-Driven Diffusion}, a framework that constructs a diffusion-like generator for spatiotemporal fields from a learned temporal predictor. At each diffusion scale $\lambda$, the predictor emulates stochastic dynamics in physical time $t$, using only present and past information to remain causal. The emulated dynamics induces a tractable surrogate path density. Its score can be computed via automatic differentiation, enabling reverse-$\lambda$ sampling for spatiotemporal fields. The framework thus unifies simulation, unconditional generation, and super-resolution using a single neural network (Fig.~\ref{fig:schematic}).

\subsection{Preparation}
\label{subsec:preparation}

We discretize the spatial coordinate $x$ on $\{x^{(i)}\}_{i=1}^{N_x}$ and physical time $t$ on $\{t^{(n)}\}_{n=0}^{N_t}$, with $t^{(0)}=0$ and $t^{(N_t)}=T$; here $N_t$ denotes the number of time steps (increments), so there are $N_t+1$ snapshots. At each $\lambda$, a spatiotemporal sample is the tensor $u_\lambda := \big(u_\lambda(x^{(i)},t^{(n)})\big)_{i,n}\in \mathbb{R}^{N_x\times (N_t+1)\times N_c}$; we write $\{u_\lambda\}_t$ for this path. For a scalar function $h(\{u_\lambda\}_t)$, the gradient $\nabla_{u_\lambda}h\in\mathbb{R}^{N_x\times (N_t+1)\times N_c}$ acts componentwise: $\nabla_{u_\lambda}h := \partial h / \partial (u_\lambda(x^{(i)},t^{(n)}))$. When clear from context, we write $u_\lambda$ for $u_\lambda(x^{(i)},t^{(n)})$. We do not consider the continuous limit; continuous-looking notation such as $\int_{\rm x, t}$ is a compact shorthand for discrete sums, following standard physics convention (\cref{subsec:notation}).

The forward RG diffusion in Eq.~(\ref{eq:forward-rg-sde}) admits a closed-form solution at each $\lambda$:
\begin{align}
    u_\lambda = \mathcal{C}_\lambda u_0 + \sqrt{\Sigma_\lambda} \epsilon, \label{eq:ou-solution-spatiotemporal}
\end{align}
where $\Sigma_\lambda$ is the spatial covariance matrix induced by the forward diffusion, $\epsilon := \epsilon(x,t)$ denotes independent standard Gaussian variables at each $(x,t)$, and $\mathcal{C}_\lambda := e^{\lambda\alpha \nabla_x^2}$ is the coarse-graining operator. On a discrete spatial grid, the Laplacian $\nabla_x^2$ becomes a matrix, so $\mathcal{C}_\lambda$ is understood as a matrix exponential. In practice, by diagonalizing the Laplacian via Fourier expansion, both $\mathcal{C}_\lambda$ and $\Sigma_\lambda$ can be computed efficiently (\cref{subsec:coarse-graining-operator-and-covariance-matrix}). Consequently, the coarse-grained field $u_\lambda$ given $u_0$ follows a Gaussian conditional density:
\begin{align}
    \{u_\lambda\}_t \mid \{u_0\}_t &\sim \prod_{n=0}^{N_t} \mathcal{N}\!\left(\mathcal{C}_\lambda u_0(\,\cdot,t^{(n)}),\Sigma_\lambda\right) \notag \\ &=: q_\lambda(\{u_\lambda\}_t \mid \{u_0\}_t). \label{eq:conditional-density}
\end{align}
This product form reflects that coarse-graining acts independently at each $t$, with $\Sigma_\lambda$ describing only the spatial correlations induced by the Laplacian.

We assume that fine-resolution spatiotemporal samples $\{u_0\}_t$ are drawn from a data distribution $q_{\rm d}$. Convolving $q_{\rm d}$ with $q_\lambda(\{u_\lambda\}_t \mid \{u_0\}_t)$ defines the marginal probability density of coarse-grained paths:
\begin{align}
    q_\lambda\left(\left\{u_\lambda\right\}_{t}\right)=\int \mathcal{D} u_0 \, q_\lambda\left(\left\{u_\lambda\right\}_{t}\mid\left\{u_0\right\}_{t}\right) q_{\rm d}\left(\left\{u_0\right\}_{t}\right), \label{eq:joint-probability-by-RG}
\end{align}
where $\mathcal{D}u_0 := \prod_{i,n} \mathrm{d}u_0(x^{(i)},t^{(n)})$ denotes the discrete path measure (\cref{subsec:path-probability-measure-via-the-path-integral-representation}). Because coarse-graining acts only on $x$ (not on $t$), the temporal correlations in $q_\lambda(\{u_\lambda\}_t)$ are inherited from $q_{\rm d}$ while preserving causality; if the Laplacian acted on $t$, it would mix future information into the present. Our construction yields a multiscale family of spatiotemporal training distributions $\{q_\lambda\}_{\lambda \geq 0}$ compatible with causal simulation.

\subsection{Governing Equation and Path Probability}
\label{subsec:governing-equation-and-path-integral}

We model the evolution in physical time $t$ at fixed $\lambda$ using a stochastic governing equation:
\begin{align}
    \partial_t u_\lambda = f_\lambda^\theta(u_\lambda) + \sigma_\lambda \xi, \label{eq:governing-eq-along-t}
\end{align}
where $f_\lambda^\theta$ is a neural predictor (also called the drift) parameterized by $\theta$ and $\xi := \xi(x,t)$ is spatiotemporal white Gaussian noise. In the discrete setting, $\xi(x,t)$ with $\mathbb{E}[\xi(x,t)\xi(x',t')] = \delta(t-t')\delta(x-x')$ corresponds to independent Gaussian variables at each grid point, with the increment $\xi(x^{(i)},t^{(n)}) \Delta t \sim \mathcal{N}(0, \Delta t)$. The stochasticity is not introduced ad hoc: spatial coarse-graining eliminates small-scale components, and temporal increments of $u_\lambda$ naturally inherit fluctuations from these eliminated degrees of freedom. The noise amplitude $\sigma_\lambda$ can then be estimated by the forward RG diffusion (\cref{subsec:specification-of-the-noise-amplitude-sigma-lambda}). In practice, the neural predictor may take a short window of past states as input \citep{Tashiro+21,Price+25}; equivalently, we augment the state so that the dynamics becomes Markov (\cref{subsec:markov-formulation-via-state-augmentation}), and develop the theory under the Markov assumption.

Under the It\^{o} convention, the stochastic dynamics in Eq.~(\ref{eq:governing-eq-along-t}) induce a tractable path density \citep{Onsager+Machlup53a,Shiraishi23}:
\begin{align}
    p_\lambda(\{u_\lambda\}_t) = \frac{r_\lambda}{Z_\lambda} \exp\!\left[- \int_{\rm x, t}\frac{\left\|\partial_t u_\lambda - f_\lambda^\theta(u_\lambda)\right\|^2}{2\sigma_\lambda^2}\right], \label{eq:path-probability-density}
\end{align}
where \(\|\cdot\|\) denotes the channel-wise Euclidean norm at each \((x,t)\), $\int_{\rm x, t} := \int \mathrm{d}t\,\mathrm{d}x\,$, $r_\lambda$ is the initial density for the spatial field $u_\lambda(\cdot, t^{(0)})$ and $Z_\lambda$ is a normalization constant. This path-integral formulation provides a convenient way to preserve causality; other representations also exist \citep[e.g.,][]{Hertz+16}. Equation~(\ref{eq:path-probability-density}) is a compact notation for the discrete-time path density arising from the Euler--Maruyama discretization, with $\int \mathrm{d}t\,\mathrm{d}x$ denoting the discrete sum over grid points (\cref{subsec:path-probability-measure-via-the-path-integral-representation}). For simplicity, we treat $r_\lambda$ as empirically given without learnable parameters (\cref{subsec:detailed-implementation}). Because $\sigma_\lambda$ is state-independent, $Z_\lambda$ depends only on $\sigma_\lambda$, $\Delta t$, and the grid dimensions, not on the learned drift $f_\lambda^\theta$. The log-density is differentiable with respect to $u_\lambda$ at each grid point, enabling the computation of the path score $\nabla_{u_\lambda}\ln p_\lambda(\{u_\lambda\}_t)$ via automatic differentiation---exactly what reverse-$\lambda$ sampling requires.

\subsection{Loss Function}
\label{subsec:loss-function}

We train the predictor $f_\lambda^\theta$ by minimizing the Kullback--Leibler (KL) divergence between the data path density $q_\lambda(\{u_\lambda\}_t)$ and the surrogate path density $p_\lambda(\{u_\lambda\}_t)$:
\begin{align}
    &D_{\mathrm{KL}}(q_\lambda \| p_\lambda) = \int \mathcal{D}u_\lambda\, q_\lambda(\{u_\lambda\}_t) \ln \frac{q_\lambda(\{u_\lambda\}_t)}{p_\lambda(\{u_\lambda\}_t)} \notag \\
    &= \frac{1}{2\sigma_\lambda^2} \mathbb{E}_{q_\lambda}\!\left[\int_{\rm x, t}\left\|\partial_t u_\lambda - f_\lambda^\theta(u_\lambda)\right\|^2\right] + \mathrm{const.} \label{eq:kl-divergence}
\end{align}
Because $Z_\lambda$ is independent of $f_\lambda^\theta$, all $\theta$-independent terms are absorbed into the constant. Minimizing the KL divergence thus reduces to a weighted regression on temporal increments. Taking $\lambda \in [0,1]$, we average over $\lambda$ to obtain the training loss:
\begin{align}
    \mathcal{L}(\theta) = \mathbb{E}_{\lambda\sim\mathcal{U}(0,1), q_\lambda}\!\left[\frac{1}{2\sigma_\lambda^2} \int_{\rm x, t}\left\|\partial_t u_\lambda - f_\lambda^\theta(u_\lambda)\right\|^2\right]. \label{eq:training-loss}
\end{align}
A single neural network $f_\lambda^\theta$ is trained across all $\lambda$, with $\lambda$ provided as an additional input. This multiscale training enables both simulation at any fixed $\lambda$ and reverse-$\lambda$ sampling for generation.

To interpret what the predictor learns, consider an idealized case where the fine-resolution (i.e., not coarse-grained) dynamics is governed by $\partial_t u_0 = f_0^{\rm true}(u_0)$. Under the joint distribution $q_\lambda(\{u_\lambda\}_t \mid \{u_0\}_t)\, q_{\rm d}(\{u_0\}_t)$, minimizing $\mathcal{L}(\theta)$ reduces to a simple regression on temporal increments, yielding an optimal drift (\cref{subsec:surrogate-model-obtained-by-kl-divergence-minimization}).
\begin{align}
    f_\lambda^*(u_\lambda) \approx \mathbb{E}\big[\mathcal{C}_\lambda f_0^{\rm true}(u_0) \mid u_\lambda\big], \label{eq:conditional-expectation-drift}
\end{align}
where the expectation is taken under the joint distribution and conditioned on $u_\lambda$ at each time step. The full expression includes an additional denoising correction that vanishes as $\lambda \to 0$; see \cref{subsec:surrogate-model-obtained-by-kl-divergence-minimization,subsec:specification-of-the-noise-amplitude-sigma-lambda} for details.

The conditional expectation in Eq.~(\ref{eq:conditional-expectation-drift}) highlights the key ordering: the true drift $f_0^{\rm true}(u_0)$ is evaluated on the fine-resolution field---including all small-scale components---and \emph{then} coarse-grained by $\mathcal{C}_\lambda$. This ordering preserves how small-scale dynamics influence large-scale evolution. In contrast, evaluating the drift on an already coarse-grained field, $f_0^{\rm true}(\mathcal{C}_\lambda u_0)$, would lose such contributions.

\subsection{Inference: Simulation and Generation}
\label{subsec:inference}

Once $f_\lambda^\theta$ is trained, inference proceeds in two modes: simulation and generation.

\paragraph{Simulation.}
We fix the diffusion scale $\lambda$, provide a coarse-grained initial condition, and integrate the governing equation~(\ref{eq:governing-eq-along-t}) forward in physical time $t$ using the learned drift $f_\lambda^\theta$. For deterministic evaluation, we set $\sigma_\lambda = 0$.

\paragraph{Generation.}
For generation, we use the reverse equation with respect to $\lambda$:
\begin{align}
    \partial_\lambda u_\lambda = \alpha \nabla_x^2 u_\lambda - \beta^2 s_\lambda + \beta \eta_\lambda, \label{eq:reverse-sde-spatiotemporal}
\end{align}
where $s_\lambda := \nabla_{u_\lambda}\ln p_\lambda(\{u_\lambda\}_t)$ is the path score and $\eta_\lambda := \eta_\lambda(x,t)$ is white Gaussian noise. After training, $p_\lambda \approx q_\lambda$; reverse-$\lambda$ sampling then generates spatiotemporal fields approximately distributed as $q_0$ ($=q_{\rm d}$). Since $p_\lambda$ in Eq.~(\ref{eq:path-probability-density}) is defined explicitly by the learned predictor, the score can be computed via automatic differentiation without training a separate score network. This avoids backpropagating through repeated evaluations of $f_\lambda^\theta$ in a physical-time rollout; we discuss the resulting computational cost in \cref{subsec:computational-cost}.

For unconditional generation, we sample a noisy path at large $\lambda$ and integrate the reverse equation toward $\lambda = 0$, yielding high-resolution spatiotemporal samples. The same procedure enables super-resolution: we start from a simulated coarse-grained path $\{u_\lambda\}_t$ at some $\lambda > 0$ and refine it toward $\lambda = 0$.

\section{Experiments}
\label{sec:experiments}

We evaluate the proposed framework on two representative chaotic systems: the 1D Lorenz-96 model with two timescales \citep{Lorenz96} and the 2D Kolmogorov flow \citep{Obukhov+83}. Both exhibit multiscale structures where small- and large-scale components are coupled. For each system, we test simulation, unconditional generation, and super-resolution using a single trained network. The goal is to validate the framework rather than to achieve state-of-the-art performance on specific benchmarks. Implementation details and extended results are provided in \cref{sec:details-of-experimental-methods,app:experimental-results,app:additional-experimental-results}. Code is available at \url{https://github.com/YukiYasuda2718/predictor-driven-diffusion}.

\subsection{Experimental Setup}
\label{subsec:experimental-setup}

\begin{figure*}[t]
    \vskip 0.0in
    \begin{center}
        \includegraphics[width=\textwidth]{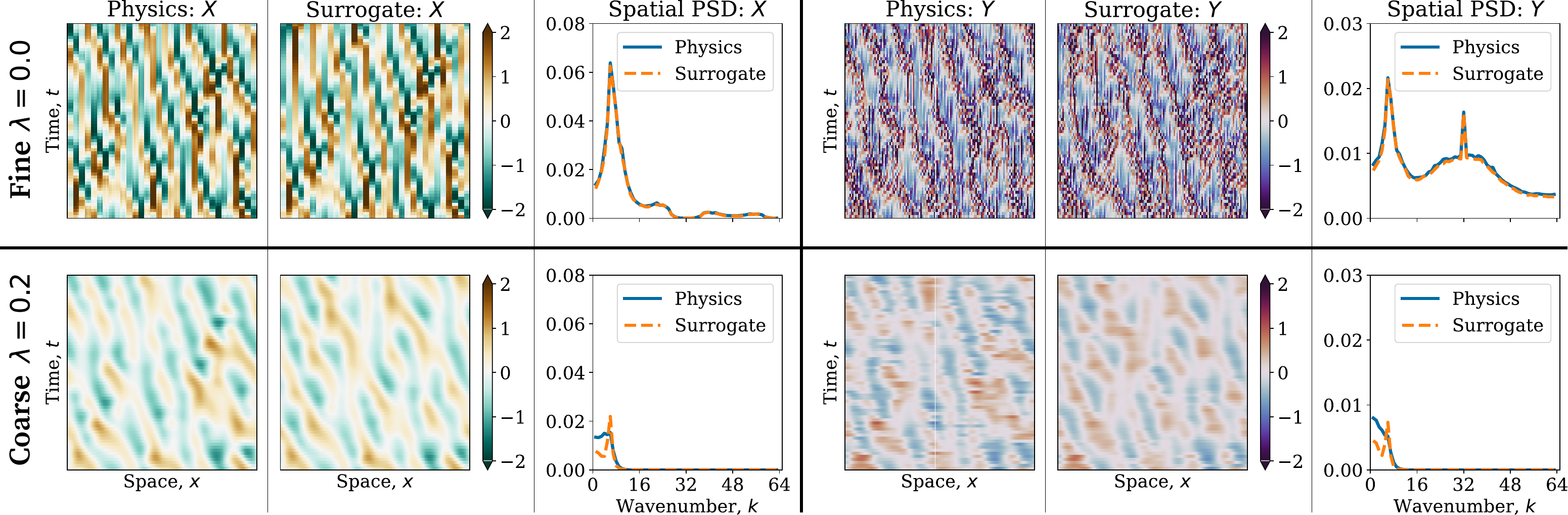}
        \caption{Simulation results for the Lorenz-96 model at fine resolution ($\lambda = 0$, top) and coarse-grained resolution ($\lambda=0.2$, bottom). Left: slow variable $X$; right: fast variable $Y$. Each panel shows spatiotemporal evolution on the left and time-averaged spatial power spectral density (PSD) on the right. The surrogate model accurately reproduces both the spatiotemporal patterns and spectral statistics of physics-based simulations. All quantities are non-dimensionalized.}
        \label{fig:main_text_lorenz96_simulation}
    \end{center}
    \vskip -0.15in
\end{figure*}

\begin{figure}[h]
    \vskip 0.1in
    \begin{center}
        \includegraphics[width=\columnwidth]{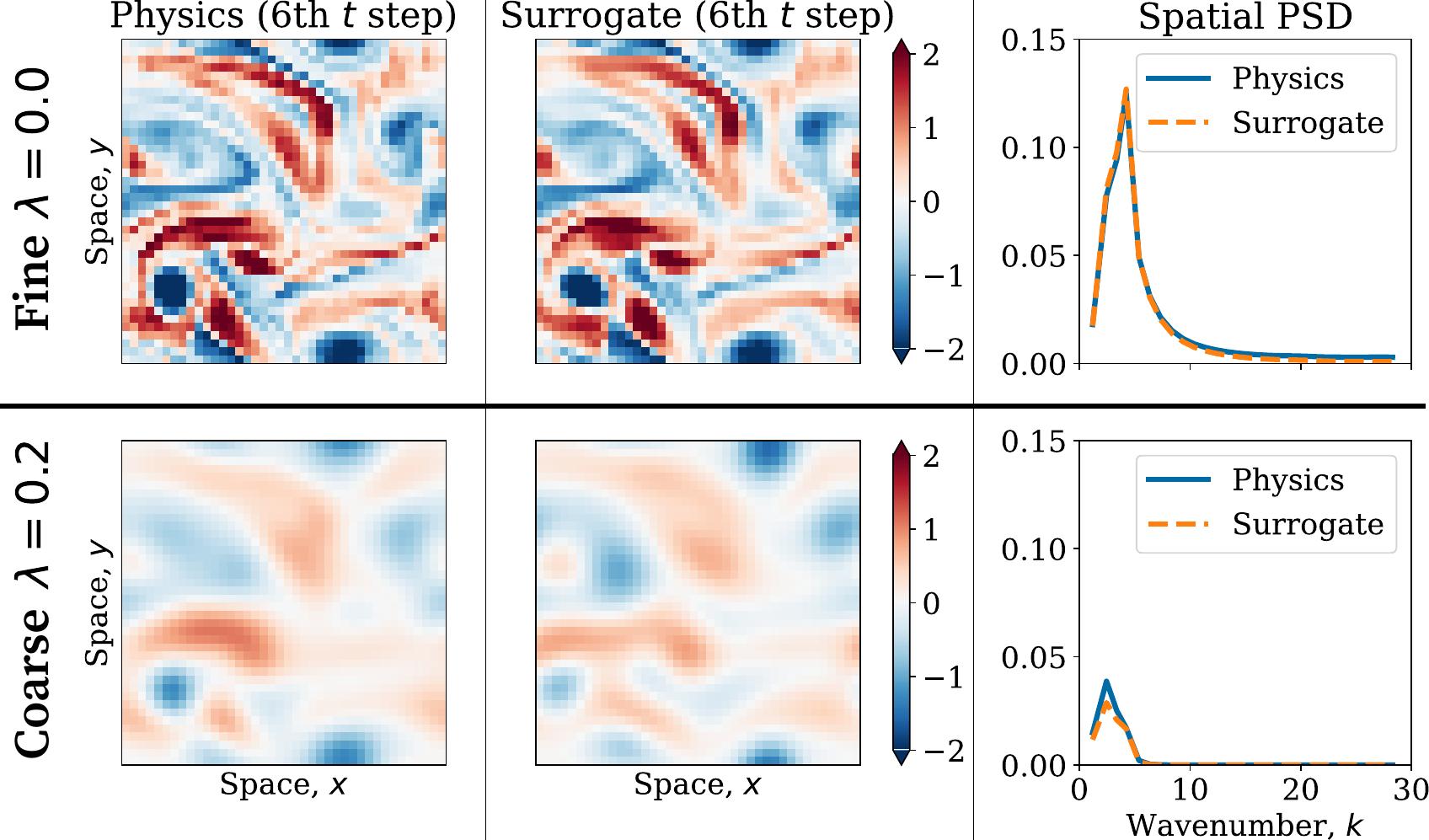}
        \caption{Simulation results for the Kolmogorov flow at fine resolution ($\lambda = 0$, top) and coarse-grained resolution ($\lambda=0.2$, bottom). Left: vorticity field at the sixth time step of the surrogate simulation; right: time-averaged spatial power spectral density (PSD). Coarse-graining removes small-scale structure while preserving large-scale dynamics. All quantities are non-dimensionalized.}
        \label{fig:main_text_kolmogorov_simulation}
    \end{center}
    \vskip -0.1in
\end{figure}

\paragraph{Training Data}
For both systems, spatiotemporal samples are generated from physics-based simulations in a statistically steady regime. Each Lorenz-96 sample has two channels $(X, Y)$ with a temporal length of 64 snapshots and a spatial length of 128 grid points. We use 3,000 and 100 samples for training and testing, respectively. Each Kolmogorov flow sample has one vorticity channel with a temporal length of 40 snapshots and a spatial resolution of $40 \times 40$. We use 6,000 and 100 samples for training and testing, respectively.

\paragraph{Neural Network}
We use a U-Net architecture \citep{Saharia+22b} for the predictor $f_\lambda^\theta$. The network takes as input the current state and the four preceding states, concatenated along the channel dimension. The output is a finite-difference approximation of the time derivative at the current time. Training minimizes the objective in Eq.~(\ref{eq:training-loss}) using local time derivatives, without rollout in $t$. We also use a Fourier Neural Operator (FNO) \citep{Li+21} to confirm that our method is not sensitive to the choice of network architecture; results are reported in \cref{subsec:experimental-results-with-fno}.

\paragraph{Hyperparameters}
RG coarse-graining is implemented analytically via the Laplacian-based operator $\mathcal{C}_\lambda$ and covariance $\Sigma_\lambda$ in Eq.~(\ref{eq:ou-solution-spatiotemporal}). The diffusion scale $\lambda \in [0,1]$ is discretized into 1,000 steps ($\lambda_{\min}=0$ and $\lambda_{\max}=1$); the parameters $\alpha$ and $\beta$ are selected via grid search: $(\alpha, \beta) = (0.1, \sqrt{2})$ for the Lorenz-96 model and $(0.3, \sqrt{6})$ for the Kolmogorov flow. In the main text, we report results at $\lambda = 0$ (fine) and $\lambda=0.2$ (coarse). In Fourier space, $\mathcal{C}_\lambda$ attenuates modes by $\exp(-\alpha\left\|k\right\|^2\lambda)$, so $\lambda=0.2$ effectively removes modes with $k_{\mathrm{cut}}\sim 1/\sqrt{\alpha\lambda}$ (about 7 for the Lorenz-96 model and about 4 for the Kolmogorov flow).

\paragraph{Baseline Models}
As a baseline, we use a denoising diffusion probabilistic model (DDPM) trained with denoising score matching \citep{Ho+20,Vincent11}, using the same U-Net architecture as our model. Separate models are trained for simulation and generation. For simulation, the baseline follows a conditional diffusion setup \citep[e.g.,][]{Price+25}, using the same temporal window as our model. Since the baseline does not define a spatial coarse-graining hierarchy, it cannot be evaluated at coarse-grained resolutions ($\lambda>0$); thus, we compare simulations only at $\lambda=0$. For generation, the baseline produces spatiotemporal samples unconditionally.

\paragraph{Evaluation Metrics}
We assess short-term predictive accuracy using the relative $L^2$ error and long-term statistical consistency using the relative spatiotemporal spectral error \citep{Jiang+23}. The relative $L^2$ error compares surrogate predictions with physics-based simulations starting from identical initial conditions. Since the physical systems are deterministic, we set $\sigma_\lambda = 0$ for simulation; see \cref{subsec:simulation-with-noise} for $\sigma_\lambda > 0$. For $\lambda > 0$, physics-based solutions are obtained by applying the coarse-graining operator $\mathcal{C}_\lambda$ to the original data $u_0$. For the spectral error, we discard the first 10 segments (each spanning the temporal length of one sample) and compute PSDs from the remaining segments; using a much longer burn-in (e.g., 100 segments) yields similar spectral errors. To estimate statistical uncertainty, we train the network five times with different random initializations and compute standard deviations. Further details are provided in \cref{subsec:evaluation-metrics}.

\subsection{Experimental Results}
\label{subsec:experimental-results}

\paragraph{Simulation}
The learned predictor accurately emulates both fine ($\lambda = 0$) and coarse-grained ($\lambda=0.2$) dynamics (Figs.~\ref{fig:main_text_lorenz96_simulation} and~\ref{fig:main_text_kolmogorov_simulation}). As $\lambda$ increases, small-scale components are progressively removed while large-scale structures are preserved. The retained large-scale components still fluctuate, and this variability is driven by the eliminated small scales; extracting only large-scale components would fail to capture such dynamics. For comparison with the DDPM baseline at $\lambda = 0$, our model shows comparable accuracy (Table~\ref{table:main_text_error_simulation}). Detailed results, such as the $\lambda$-dependence of prediction accuracy, are provided in \cref{subsec:details-of-experimental-results-simulation}.

\begin{table}[t]
    \vskip 0.09in
    \caption{Simulation errors at $\lambda = 0$ for the Lorenz-96 model (L96) and Kolmogorov flow (KF), comparing our predictor-driven approach with the DDPM baseline. $L^2$ error measures short-term prediction accuracy (at the sixth time step); spectral error measures long-run statistical consistency. Values are means $\pm$ standard deviations across five training runs (100 test samples each). Bold indicates better performance. Both methods achieve comparable accuracy.}
    \label{table:main_text_error_simulation}
    \centering
    \small
    \scshape
    \resizebox{\columnwidth}{!}{
        \begin{tabular}{lcc}
            \toprule
                            & Predictor-Driven             & DDPM (Baseline)               \\
            \midrule
            L96 $L^2$ Error & \textbf{0.503} ($\pm$ 0.011) & 0.557 ($\pm$ 0.031)           \\
            KF $L^2$ Error  & \textbf{0.691} ($\pm$ 0.057) & 0.861 ($\pm$ 0.023)           \\
            \midrule
            L96 Spec. Error & \textbf{0.120} ($\pm$ 0.014) & 0.226 ($\pm$ 0.077)           \\
            KF Spec. Error  & 0.182 ($\pm$ 0.029)          & \textbf{0.139} ($\pm$ 0.021)  \\
            \bottomrule
        \end{tabular}
    }
    \vskip 0.08in
\end{table}

\paragraph{Generation}
Using the same network trained for simulation, we generate spatiotemporal samples by integrating the reverse equation~(\ref{eq:reverse-sde-spatiotemporal}) in $\lambda$ starting from Gaussian noise. This validates that the predictor-defined path density supplies a usable score for diffusion-like generation. Table~\ref{table:main_text_error_generation} compares spectral errors with the DDPM baseline; our method achieves comparable performance, demonstrating the effectiveness of the unified framework beyond simulation. A detailed comparison is provided in \cref{subsec:details-of-experimental-results-spatiotemporal-data-generation}.

\newpage

\begin{table}[t]
    \caption{Spectral error for unconditional generation on the Lorenz-96 model (L96) and Kolmogorov flow (KF), comparing our predictor-driven approach with the DDPM baseline. Spectral error measures statistical consistency of spatiotemporal patterns between generated and physics-based samples. Values are means $\pm$ standard deviations across five training runs (100 test samples each). Bold indicates better performance. All quantities are non-dimensionalized.}
    \label{table:main_text_error_generation}
    \centering
    \small
    \scshape
    \resizebox{\columnwidth}{!}{
        \begin{tabular}{lcc}
            \toprule
                            & Predictor-Driven              & DDPM (Baseline)              \\
            \midrule
            L96 Spec. Error & 0.343 ($\pm$ 0.028)           & \textbf{0.267} ($\pm$ 0.020) \\
            KF Spec. Error  & \textbf{0.457} ($\pm$ 0.025)  & 0.615 ($\pm$ 0.121)          \\
            \bottomrule
        \end{tabular}
    }
    \vskip -0.1in
\end{table}

\begin{figure}[h]
    \vskip 0.175in
    \begin{center}
        \includegraphics[width=\columnwidth]{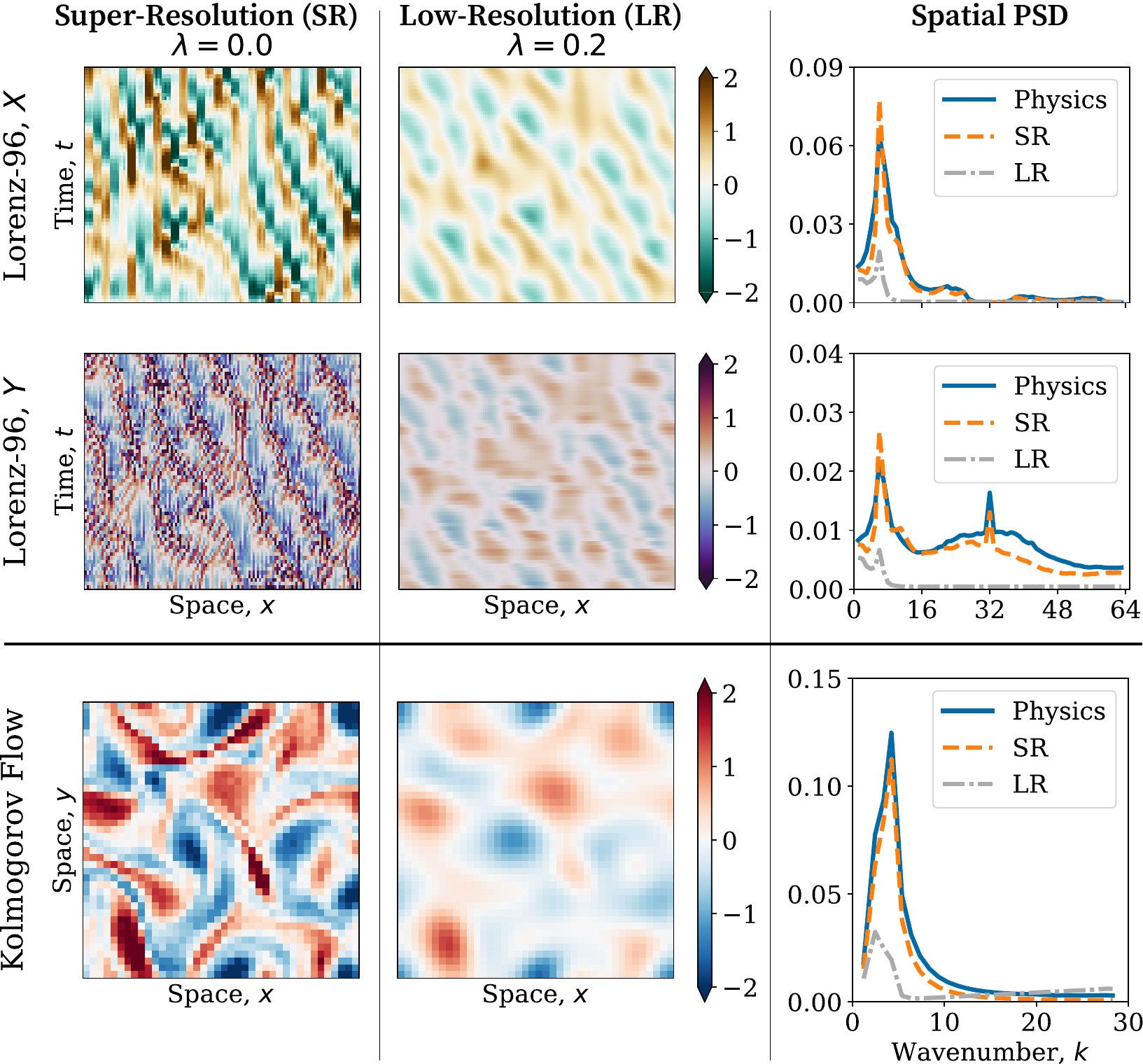}
        \caption{Super-resolution results for the Lorenz-96 model (top) and Kolmogorov flow (bottom), shown at the sixth time step. Left: super-resolved output at $\lambda = 0$; middle: low-resolution input from surrogate simulation at $\lambda=0.2$; right: time-averaged spatial power spectral densities (PSD; computed from 100 samples) comparing low-resolution input, super-resolved output, and physics-based simulation. The reverse-$\lambda$ integration restores small-scale structure absent from the low-resolution input. All quantities are non-dimensionalized.}
        \label{fig:main_text_super_resolution}
    \end{center}
\end{figure}

\paragraph{Super-Resolution}
Super-resolution is achieved by treating a coarse-grained simulation at $\lambda=0.2$ as input and refining it to $\lambda = 0$ via reverse-$\lambda$ integration (Fig.~\ref{fig:main_text_super_resolution}). This procedure restores small-scale components absent from the low-resolution input. The restored fine structure exhibits spectral statistics comparable to those of unconditional generation (Tables~\ref{table:main_text_error_generation} and~\ref{table:main_text_error_super_resolution}). These results demonstrate that simulation and super-resolution are unified through one trained model.

\section{Related Work}

\paragraph{Diffusion and Flow-Based Generative Models}

Generative models based on stochastic and ordinary differential equations (SDEs and ODEs) have been extensively studied \citep{Song+21,Lipman+23,Albergo+23}. These models treat a diffusion scale $\lambda$ as a time-like parameter---often called diffusion time (or step)---with a neural network modeling the score or drift. In contrast, our approach models the drift in physical time $t$. This distinction affects the training objective: standard diffusion models minimize KL divergences defined in diffusion time \citep{Ho+20,Kingma+21}, whereas our model minimizes a KL divergence between data and surrogate path densities defined in physical time.

In many generative models, intermediate representations along the diffusion-time axis are not explicitly interpreted, as their semantic meaning is generally unclear. Consequently, recent work has focused on reducing the number of integration steps by tuning sampling strategies or noise schedules \citep{Nichol+21,Karras+22}, or by straightening the transformation trajectory \citep{Liu+23,Tong+23}. In contrast, our framework employs RG-based diffusion processes, which provide a clear interpretation of intermediate representations: each $\lambda$ corresponds to dynamics coarse-grained at a specific spatial scale.

\begin{table}[t]
    \caption{Spectral error for super-resolution on the Lorenz-96 model (L96) and Kolmogorov flow (KF). SR: super-resolved output at $\lambda = 0$ from low-resolution input at $\lambda=0.2$; LR: low-resolution input evaluated directly against the $\lambda = 0$ physics-based simulation. Super-resolution substantially reduces spectral error compared to the low-resolution input, demonstrating effective reconstruction of small-scale structure. Values are means $\pm$ standard deviations across five training runs (100 test samples each).}
    \label{table:main_text_error_super_resolution}
    \centering
    \small
    \scshape
    \resizebox{\columnwidth}{!}{
        \begin{tabular}{lcc}
            \toprule
                            & SR & LR at $\lambda=0.2$ (vs.\ $\lambda=0$) \\
            \midrule
            L96 Spec. Error & \textbf{0.345} ($\pm$ 0.027)  & 0.873 ($\pm$ 0.005)    \\
            KF Spec. Error  & \textbf{0.303} ($\pm$ 0.011)  & 0.743 ($\pm$ 0.005)    \\
            \bottomrule
        \end{tabular}
    }
    \vskip 0.05in
\end{table}

\paragraph{RG-Based Diffusion Models}

RG-based diffusion models have been studied for static data such as images \citep{Cotler+Rezchikov23,Sheshmani+25,Masuki+Ashida25}. These models employ damping and noise levels that depend on both wavenumber $k$ and diffusion scale $\lambda$; the Laplacian-based formulation we use is one instance of this broader class. These approaches improve computational efficiency and accuracy over standard diffusion \citep{Sheshmani+25,Masuki+Ashida25}. Since we do not optimize damping and noise schedules, our model could similarly adopt $k$- and $\lambda$-dependent damping and noise levels to improve efficiency or accuracy, which we leave to future work.

\paragraph{Other Scale-Aware Diffusion Models}

Several diffusion models introduce multiscale structure without modifying the forward process. Some use hierarchical networks or multiresolution representations \citep{Ryu+Ye22,Fan+23,Phung+23}, while others factorize the data distribution across scales and estimate conditional scores \citep{Guth+22}. RG-based diffusion differs fundamentally by incorporating scale dependence in the forward process through explicit coarse-graining. Despite this distinction, both paradigms share the principle of coarse-to-fine generation \citep[e.g.,][]{Guth+22}, suggesting potential avenues for integration.

\paragraph{Neural ODEs and SDEs}

Neural networks have been used to approximate ODEs or SDEs in physical time $t$, typically at a fixed resolution without an explicit diffusion scale $\lambda$ \citep{Legaard+23,North+23}. For dynamics prediction, some methods use diffusion time as a proxy for physical time \citep{Ruhling+23,Ruhling+24}. For generation, others introduce dynamical-system perspectives into continuous-time neural models \citep{Chen+18,Kidger+21}. Our method explicitly distinguishes physical time $t$ from the diffusion scale $\lambda$, learning a $\lambda$-conditioned family of predictors tied to spatial coarse-graining, not a single-resolution surrogate.

\paragraph{Energy-Based Generative Models}

Energy-based models define a probability density as $p(x) \propto \exp(-E(x))$ with a learned energy $E(x)$ \citep{Gao+21,Zhu+24}. These models use differentiable log-densities to support both sampling and learning. Our path-probability formulation plays a formally analogous role: the exponent in Eq.~(\ref{eq:path-probability-density}) serves as a path energy, enabling the computation of path-wise scores for reverse-$\lambda$ sampling. While energy-based approaches primarily target static data distributions, \citet{Shen+Chen25} proposed a likelihood-based framework for temporal dynamics, albeit without explicitly addressing multiscale structure. Our method integrates temporal dynamics with explicit spatial multiscale modeling through RG coarse-graining.

\paragraph{Super-Resolution}

Diffusion models have been widely applied to super-resolution \citep{Ho+22b,Saharia+22b,Saharia+22a,Mardani+25}. 
Some approaches interpret or exploit intermediate diffusion states as coarse-grained representations, and perform super-resolution by integrating from these states along the diffusion-scale axis \citep{Shu+23,Bischoff+Deck24}. When combined with prediction tasks, a common strategy is to perform low-resolution surrogate prediction, followed by super-resolution \citep{Chen+23,Brenowitz+25}, but the predictor and super-resolver are typically trained separately. Our framework unifies these steps: the predictor defines a path model whose score enables reverse-$\lambda$ super-resolution.

\section{Conclusions}
\label{sec:conclusions}

We proposed Predictor-Driven Diffusion, a framework that combines RG-based spatial coarse-graining in diffusion scale $\lambda$ with a path-integral formulation of temporal evolution in physical time $t$. This construction gives intermediate diffusion states a clear interpretation as a spatial hierarchy of coarse-grained dynamics. The key idea is to train a predictor on the KL divergence between path densities across a range of $\lambda$ to capture multiscale dynamics. This two-axis construction enables forward-$t$ simulation and reverse-$\lambda$ sampling, unifying simulation, generation, and super-resolution in a single neural network. We validated the framework on the Lorenz-96 model and Kolmogorov flow, two representative multiscale dynamical systems.

\paragraph{Limitations}

Our framework targets spatiotemporal fields governed by physical dynamics; its applicability to general video data without explicit governing equations remains unexplored. We treat the initial density $r_\lambda$ as empirically given, which is appropriate for stationary systems but requires extension for transient settings. We validated the framework only on the 1D Lorenz-96 model and 2D Kolmogorov flow; scalability to higher resolutions and three-dimensional data remains untested. Generation and super-resolution are more costly than simulation due to path-score computation at each reverse-$\lambda$ step. The coarse-graining parameters $\alpha$ and $\beta$ may require system-specific tuning; adopting $\lambda$-dependent $\alpha$ (damping) and $\beta$ (noise amplitude) may further improve accuracy.

\section*{Acknowledgements}
The deep learning experiments were performed on the Earth Simulator at the Japan Agency for Marine-Earth Science and Technology (JAMSTEC).

\section*{Impact Statement}

This paper presents work whose goal is to advance the field of machine learning. By connecting renormalization group theory from statistical physics with diffusion-based generative models, our work offers a new perspective on surrogate modeling for multiscale dynamical systems. We believe this interdisciplinary approach may contribute to both fields and could facilitate more efficient scientific simulations, such as climate modeling. Although our study primarily addresses methodological aspects, we recognize the potential societal implications of AI advancements and emphasize the importance of validating surrogate models against physics-based simulations before deployment in safety-critical applications.

\bibliography{references}
\bibliographystyle{icml2026}

\newpage
\appendix
\crefalias{section}{appendix}
\crefalias{subsection}{appendix}
\onecolumn

\section{Theoretical Details}
\label{app:theoretical-details}

\subsection{Notation}
\label{subsec:notation}

We adopt spatiotemporal discretization throughout this paper, a standard convention in statistical physics \citep[e.g.,][]{Shiraishi23} that has also been used in diffusion-model research \citep{Hirono+24}. We do not consider the continuous limit ($\Delta t \to 0$). We summarize the notation and conventions below.

\paragraph{Spatiotemporal discretization}
We adopt the It\^{o} convention and treat the time increment $\Delta t$ as sufficiently small but fixed. Physical time $t$ is discretized as $\{t^{(n)}\}_{n=0}^{N_t}$ with $t^{(n)} = n\Delta t$, and the spatial coordinate $x$ is discretized as $\{x^{(i)}\}_{i=1}^{N_x}$. Each spatiotemporal sample at diffusion scale $\lambda$ is a tensor $u_\lambda \in \mathbb{R}^{N_x \times (N_t+1) \times N_c}$, where $N_c$ denotes the number of field channels.

All computations are performed in discrete $x$ and $t$; continuous notation such as $\int_{\rm x, t}$, $\int \mathrm{d}t\,\mathrm{d}x$, and $\int \mathrm{d}t$ is a compact shorthand. For instance, the integral symbol $\int_{\rm x, t}$ used in the main text denotes $\int_{\rm x, t} = \int \mathrm{d}x\,\mathrm{d}t = \sum_{i,n} \Delta x \, \Delta t$. As for derivatives, we follow the Euler--Maruyama discretization: $(\partial_t u_\lambda)(t^{(n)}) = (u_\lambda(t^{(n+1)}) - u_\lambda(t^{(n)}))/\Delta t$, where we omit the spatial argument $x^{(i)}$.

\paragraph{Path notation}
We write $\{u_\lambda\}_t$ for the spatiotemporal path, i.e., the collection of field values $\big(u_\lambda(x^{(i)}, t^{(n)})\big)_{i,n}$ over all grid points. The discrete path measure is denoted by $\mathcal{D}u_\lambda := \prod_{i,n} \mathrm{d}u_\lambda(x^{(i)}, t^{(n)})$. For any scalar function $h(\{u_\lambda\}_t): \mathbb{R}^{N_x \times (N_t+1) \times N_c} \to \mathbb{R}$, the gradient $\nabla_{u_\lambda}h \in \mathbb{R}^{N_x \times (N_t+1) \times N_c}$ acts componentwise: $\nabla_{u_\lambda}h := \partial h / \partial (u_\lambda(x^{(i)},t^{(n)}))$.

\paragraph{Probability densities}
We denote probability densities by lowercase letters: $p$, $q$, and $r$. These are densities, not probability measures (i.e., they do not include the differential $\mathcal{D}u$). Specifically, $q_{\rm d}(\{u_0\}_t)$ denotes the data path density at $\lambda = 0$; $q_\lambda(\{u_\lambda\}_t)$ denotes the coarse-grained data path density at diffusion scale $\lambda$ in Eq.~(\ref{eq:joint-probability-by-RG}); $q_\lambda^{\rm joint}(\{u_\lambda\}_t, \{u_0\}_t)$ denotes the joint density of $\{u_\lambda\}_t$ and $\{u_0\}_t$ (\cref{subsec:surrogate-model-obtained-by-kl-divergence-minimization}); $p_\lambda(\{u_\lambda\}_t)$ denotes the surrogate path density induced by the learned predictor $f_\lambda^\theta$ in Eq.~(\ref{eq:path-probability-density}); and $r_\lambda$ denotes the initial density for $u_\lambda(\,\cdot\,, t^{(0)})$.

\paragraph{Scales and operators}
The diffusion scale $\lambda \geq 0$ parameterizes the degree of spatial coarse-graining and corresponds (up to constants) to the squared characteristic wavelength (\cref{subsec:rg-transformation-via-diffusion-processes} and \cref{subsec:coarse-graining-operator-and-covariance-matrix}). In Eqs.~(\ref{eq:ou-solution-spatiotemporal}) and~(\ref{eq:conditional-density}), the coarse-graining operator $\mathcal{C}_\lambda := e^{\lambda \alpha \nabla_x^2}$ acts only on spatial coordinates, and the spatial covariance matrix induced by the forward diffusion is denoted by $\Sigma_\lambda$. The noise amplitude in the governing equation~(\ref{eq:governing-eq-along-t}) is $\sigma_\lambda$.

\paragraph{White Gaussian noise}
We use two white Gaussian noise variables: $\eta_\lambda := \eta_\lambda(x,t)$ for the diffusion processes in Eqs.~(\ref{eq:forward-rg-sde}) and~(\ref{eq:reverse-sde-spatiotemporal}), and $\xi := \xi(x,t)$ for the physical-time dynamics in Eq.~(\ref{eq:governing-eq-along-t}). Both are independent and identically distributed (i.i.d.) across all grid points.

The noise $\eta_\lambda$ is sampled at each diffusion scale $\lambda$: in discrete form, $\eta_\lambda \sim \mathcal{N}(0, \frac{1}{\Delta \lambda} I_{D'})$, so that the increment $\eta_\lambda \Delta\lambda \sim \mathcal{N}(0, \Delta\lambda \, I_{D'})$. Here $D' := N_x \times (N_t+1) \times N_c$ denotes the total spatiotemporal degrees of freedom. This $1/\Delta \lambda$ scaling ensures that the noise increments have variance proportional to the step size, consistent with the standard Wiener process convention \citep[e.g.,][]{Shiraishi23}.

The noise $\xi$ is sampled at each time step $t^{(n)}$: in discrete form, $\xi(t^{(n)}) \sim \mathcal{N}(0, \frac{1}{\Delta t} I_D)$, so that the increment $\xi(t^{(n)}) \Delta t \sim \mathcal{N}(0, \Delta t \, I_D)$. Here $D := N_x \times N_c$ denotes the number of spatial degrees of freedom at a fixed time step. The $1/\Delta t$ scaling similarly ensures consistency with the Wiener process convention.

\subsection{Coarse-Graining Operator and Covariance Matrix}
\label{subsec:coarse-graining-operator-and-covariance-matrix}

We derive explicit expressions for the coarse-graining operator $\mathcal{C}_\lambda$ and the covariance matrix $\Sigma_\lambda$ appearing in the forward solution~(\ref{eq:ou-solution-spatiotemporal}) and the conditional density~(\ref{eq:conditional-density}). To obtain diagonal representations in Fourier space, we consider a finite spatial domain with periodic boundary conditions, $x \in [0, 2\pi)^d$.
\begin{align}
    \widetilde{u}_\lambda(k,t) &= \mathcal{F}[u_\lambda(x,t)] = \int_{[0, 2\pi)^d} \mathrm{d}x \frac{e^{-i k \cdot x}}{(\sqrt{2\pi})^d} u_\lambda(x,t),  \\
    {u}_\lambda(x,t) &= \mathcal{F}^{-1}[\widetilde{u}_\lambda(k,t)] = \sum_k \frac{e^{i k \cdot x}}{(\sqrt{2\pi})^d} \widetilde{u}_\lambda(k,t),
\end{align}
where the wavenumber is given by $k \in \mathbb{Z}^d$.

In Fourier space, the forward process~(\ref{eq:forward-rg-sde}) takes the form
\begin{align}
    \partial_\lambda \widetilde{u}_\lambda &= -\alpha \left\| k \right\|^2 \widetilde{u}_\lambda +\beta \widetilde{\eta}_\lambda.
\end{align}
Each Fourier mode evolves independently. For $k \ne 0$, it follows an Ornstein--Uhlenbeck process, for which an analytical solution can be obtained \citep[e.g.,][]{Gardiner09}:
\begin{align}
    \widetilde{u}_\lambda = e^{-\alpha \left\| k \right\|^2 \lambda} \widetilde{u}_0 + \beta \int_0^\lambda e^{-\alpha \left\| k \right\|^2\left(\lambda - \lambda'\right)} \widetilde{\eta}_{\lambda'} \mathrm{d}\lambda'.
\end{align}
The conditional density at each $t$ is therefore given by
\begin{align}
    q_\lambda(\widetilde{u}_\lambda(\cdot,t)\mid\widetilde{u}_0(\cdot,t)) \propto \prod_{k \ne 0} \exp \left(-\frac{1}{2}\frac{2 \alpha\left\| k \right\|^2}{\beta^2\left(1-e^{-2\alpha \left\| k \right\|^2 \lambda}\right)}\left|\widetilde{u}_\lambda(k,t)-\widetilde{u}_0(k,t)\, e^{-\alpha \left\| k \right\|^2 \lambda}\right|^2\right).
\end{align}
The conditional density used in the main text, $q_\lambda(\{u_\lambda\}_t \mid \{u_0\}_t)$ in Eq.~(\ref{eq:conditional-density}), is obtained by expressing this Gaussian density in real space. In Fourier space, both the coarse-graining operator and the covariance matrix are diagonal, and are given for each $k$ by
\begin{align}
    \widetilde{\mathcal{C}}_\lambda(k) &= e^{-\alpha \left\| k \right\|^2 \lambda}, \label{eq:coarse-graining-operator-in-fourier} \\
    \widetilde{\Sigma}_\lambda(k) &=\beta^2\frac{1-e^{-2\alpha \left\| k \right\|^2 \lambda}} {2\alpha \left\| k \right\|^2}.\label{eq:coarse-graining-covariance-in-fourier}
\end{align}
In practice, $\mathcal{C}_\lambda$ can be applied efficiently by (i) transforming the field $u_\lambda(x,t)$ to Fourier space, (ii) multiplying by the diagonal coefficients $\widetilde{\mathcal{C}}_\lambda(k)$, and (iii) transforming back to real space; the same procedure applies to $\Sigma_\lambda$ and $\epsilon$ in Eq.~(\ref{eq:ou-solution-spatiotemporal}).

This Fourier representation provides a direct interpretation of the diffusion scale $\lambda$ in terms of spatial scales \citep{Carosso20,Cotler+Rezchikov23}. Since the wavenumber $k\in\mathbb{Z}^d$ on the periodic domain $[0,2\pi)^d$ is dimensionless, the Fourier mode at wavenumber $k$ has spatial wavelength $2\pi/\left\|k\right\|$. Writing the wavelength as $\ell_k:=2\pi/\left\|k\right\|$, the exponent of $\widetilde{\mathcal{C}}_\lambda$ in Eq.~(\ref{eq:coarse-graining-operator-in-fourier}) can be rewritten as $\alpha\left\|k\right\|^2\lambda=\alpha\lambda(2\pi/\ell_k)^2$; hence $\sqrt{\alpha\lambda}$ corresponds to a wavelength (i.e., length scale) up to the factor $2\pi$. Equivalently, modes with $\left\|k\right\|\gg k_{\mathrm{cut}}:=1/\sqrt{\alpha\lambda}$ are exponentially suppressed (i.e., effectively have near-zero amplitude), while modes with $\left\|k\right\|\lesssim k_{\mathrm{cut}}$ remain.

Throughout the paper, we exclude the zero wavenumber mode ($k=0$), consistent with our data preprocessing (\cref{subsec:construction-of-training-data}): all variables are standardized to have zero mean, which removes the constant Fourier mode. Accordingly, products over Fourier modes should be understood as taken over $k \ne 0$. If the $k=0$ mode is retained, one can avoid the $k=0$ singularity by introducing a scale-independent damping term $-\gamma u_\lambda$ in the forward equation~(\ref{eq:forward-sde-with-gamma}) following \citet{Cotler+Rezchikov23}; we study this variant in \cref{subsec:effect-of-the-laplacian}.

In the following, we use the real-space representation to remain consistent with the main text.

\subsection{Carosso Renormalization Group}
\label{subsec:carosso-renormalization-group}

We briefly review the renormalization group (RG) proposed by Carosso and the underlying perspective of effective field theory \citep{Carosso20,Cotler+Rezchikov23,Mccomb03,Schwenk+12}. The RG transformation is defined as
\begin{align}
    q(\phi_\lambda) &:= \int \mathcal{D}\phi_0 \, q(\phi_\lambda|\phi_0) q(\phi_0), \\
    q(\phi_0) &:= \frac{1}{\mathcal{Z}_0} e^{-S_0(\phi_0)}, \\
    \mathcal{Z}_0 &:= \int \mathcal{D}\phi_0 \, e^{-S_0(\phi_0)}.
\end{align}
\citet{Carosso20} considers renormalization of a field $\phi_0(x)$ defined on spatial coordinates $x$, without incorporating physical time $t$. To be consistent with the notation commonly used in statistical field theory \citep[e.g.,][]{Mccomb03,Schwenk+12}, we adopt the field variables $\phi_0$ and $\phi_\lambda$ here. Since integration over the field $\phi_0$ is required, we introduce the measure $\mathcal{D}\phi_0 := \prod_x \mathrm{d}\phi_0(x)$ under an appropriate discretization. The probability density of the unrenormalized (i.e., bare) field $\phi_0$ is assumed to be known and is denoted by $q(\phi_0)$. In physics, probability densities are often taken to belong to the exponential family, and the exponent is referred to as the action $S_0$. This action determines the probability density of $\phi_0$ and contains parameters of interest associated with $\phi_0$, such as interaction coefficients. By changing the scale of interest through coarse-graining, that is, the RG transformation, one can track how these parameters evolve \citep[e.g.,][]{Mccomb03,Schwenk+12}.

In the \citet{Carosso20} formulation, the RG transformation is defined as a convolution with a Gaussian kernel $q(\phi_\lambda \mid \phi_0)$. This operation constitutes a coarse-graining transformation with a semigroup structure with respect to the parameter $\lambda$, thereby defining an RG transformation. Through this transformation, the mean of the probability density is damped by the operator $\mathcal{C}_\lambda$. Equivalently, small-scale components, corresponding to Fourier modes with large wavenumbers $k$, are preferentially suppressed. At the same time, diffusion induced by noise leads to a convolution of probability densities, and its effect is reflected in the statistical properties of remaining large-scale components through the effective action \citep{Carosso20,Cotler+Rezchikov23}. This formulation makes clear that both the Laplacian and the noise term are essential for the RG transformation.

Through the RG transformation, one can derive an approximate theory for the coarse-grained state, namely an effective field theory \citep{Cotler+Rezchikov23}. Based on the convolution with the Gaussian kernel, the effective action $S_\lambda$ is defined as
\begin{align}
    \int \mathcal{D}\phi_0 \, q(\phi_\lambda|\phi_0) q(\phi_0) = q(\phi_\lambda) = \frac{1}{\mathcal{Z}_0} e^{-S_\lambda(\phi_\lambda)},
\end{align}
where the normalization constant $\mathcal{Z}_0$ remains unchanged under coarse-graining \citep{Carosso20},
\begin{align}
    \mathcal{Z}_0 = \int \mathcal{D}\phi_0 \, e^{-S_0(\phi_0)} = \int \mathcal{D}\phi_\lambda\mathcal{D}\phi_0 \, q(\phi_\lambda|\phi_0) e^{-S_0(\phi_0)}.
\end{align}
That is, the entire effect of coarse-graining is absorbed into a change of the action. By examining how the original action $S_0$ is transformed into the effective action $S_\lambda$, one can track the evolution of parameters of interest contained in $S_0$, such as interaction coefficients, and thereby analyze the multiscale structure of the system \citep[e.g.,][]{Mccomb03,Schwenk+12}.

\subsection{Path Probability Density via the Path-Integral Representation}
\label{subsec:path-probability-measure-via-the-path-integral-representation}

We derive the path probability density using a path-integral representation, assuming that the dynamics in physical time $t$ are Markov \citep{Onsager+Machlup53a,Shiraishi23}. We assume that time $t$ is discretized as $t^{(n)}=n\Delta t$ with $n=0,\ldots,N_t$ (i.e., $T=N_t\Delta t$) and adopt the It\^{o} convention, under which the governing equation takes the form
\begin{align}
    u_\lambda(t^{(n+1)}) = u_\lambda(t^{(n)}) + f_\lambda^\theta(u_\lambda(t^{(n)})) \Delta t + \sigma_\lambda\xi(t^{(n)}) \Delta t. \label{eq:discrete-time-governing-equation}
\end{align}
The spatial dependence on $x$ is notationally omitted but understood to be present. Let $D$ denote the dimension of the discretized state $u_\lambda(t^{(n)})$ at a fixed time step (including spatial degrees of freedom and channels).

The white Gaussian noise follows the Gaussian distribution \citep[e.g.,][]{Shiraishi23}:
\begin{align}
    \xi(t^{(n)}) &\sim \mathcal{N}(0, \frac{1}{\Delta t} I_D). \label{eq:discrete-time-noise-probability-density}
\end{align}
Accordingly, the probability density of the noise is given by
\begin{align}
    \left(\frac{\Delta t}{2\pi}\right)^{D/2} \exp\left[-\frac{\Delta t}{2}\left\| \xi(t^{(n)}) \right\|^2 \right],
\end{align}
where $\|\cdot\|$ denotes the Euclidean norm over all spatial degrees of freedom and channels at a fixed time step; this is equivalent to integrating the channel-wise norm over space in the main-text notation (up to the $\Delta x$ discretization factor). Using the governing equation~(\ref{eq:discrete-time-governing-equation}), the probability density of $u_\lambda(t^{(n+1)})$ is derived \citep[e.g.,][]{Shiraishi23}:
\begin{align}
    p(u_\lambda(t^{(n+1)}) \mid u_\lambda(t^{(n)})) = (2\pi \sigma_\lambda^2\Delta t)^{-D/2} \exp\left[-\frac{1}{2\sigma_\lambda^2\Delta t}\left\| u_\lambda(t^{(n+1)}) - u_\lambda(t^{(n)}) - f_\lambda^\theta(u_\lambda(t^{(n)})) \Delta t \right\|^2 \right], \label{eq:transition-probability-density}
\end{align}
where we use the fact that $\sigma_\lambda$ is a state-independent constant, so that the Jacobian associated with the change of variables is given by $(\sigma_\lambda \Delta t)^{-D}$. Clearly, the normalization constant does not depend on $f_\lambda^\theta$.

The path-integral representation is obtained by taking the product of this transition probability density over successive time steps,
\begin{align}
    p_\lambda(\{u_\lambda\}_t) &= \frac{r_\lambda(u_\lambda(t^{(0)}))}{(2\pi \sigma_\lambda^2 \Delta t)^{D N_t/2}} \prod_{n=0}^{N_t-1} \exp\left[-\frac{1}{2\sigma_\lambda^2\Delta t}\left\| u_\lambda(t^{(n+1)}) - u_\lambda(t^{(n)}) - f_\lambda^\theta(u_\lambda(t^{(n)})) \Delta t \right\|^2 \right], \\
    &= \frac{r_\lambda(u_\lambda(t^{(0)}))}{Z_\lambda} \prod_{n=0}^{N_t-1} \exp\left[-\frac{\Delta t}{2\sigma_\lambda^2}\left\| \frac{u_\lambda(t^{(n+1)}) - u_\lambda(t^{(n)})}{\Delta t} - f_\lambda^\theta(u_\lambda(t^{(n)})) \right\|^2 \right], \\
    &= \frac{r_\lambda(u_\lambda(t^{(0)}))}{Z_\lambda} \exp \left[ - \frac{1}{2 \sigma_\lambda^2} \int \mathrm{d}t \left\| \partial_t u_\lambda - f_\lambda^\theta(u_\lambda) \right\|^2\right], \label{eq:path-integral-formal-representation}
\end{align}
where $r_\lambda(u_\lambda(t^{(0)}))$ denotes the probability density of the initial condition and the normalization constant is given by $Z_\lambda := (2\pi \sigma_\lambda^2 \Delta t)^{D N_t/2}$.

We regard the following quantity as an effective action $S_\lambda$ (\cref{subsec:carosso-renormalization-group}) and train $f_\lambda^\theta$ to represent the temporal evolution associated with the RG transformation,
\begin{align}
    S_\lambda = \frac{1}{2 \sigma_\lambda^2} \int\mathrm{d}t \| \partial_t u_\lambda - f_\lambda^\theta(u_\lambda) \|^2.
\end{align}
If a true physical model exists at $\lambda = 0$, minimizing the KL divergence objective in Eq.~(\ref{eq:training-loss}) leads the surrogate model to reproduce the coarse-grained dynamics (\cref{subsec:surrogate-model-obtained-by-kl-divergence-minimization}).

As $\sigma_\lambda \to 0$, the transition probability in Eq.~(\ref{eq:transition-probability-density}) converges to a delta distribution $\delta(u_\lambda(t^{(n+1)}) - u_\lambda(t^{(n)}) - f_\lambda^\theta \Delta t)$, and consequently the path density $p_\lambda(\{u_\lambda\}_t)$ concentrates on deterministic paths satisfying $\partial_t u_\lambda = f_\lambda^\theta$. Although this limit is well defined in the distributional sense, the path score $\nabla_{u_\lambda} \ln p_\lambda$ diverges due to the $1/\sigma_\lambda^2$ factor in Eq.~(\ref{eq:path-integral-formal-representation}). The numerical treatment of this limit is discussed in \cref{subsec:specification-of-the-noise-amplitude-sigma-lambda}.

\subsection{Markov Formulation via State Augmentation}
\label{subsec:markov-formulation-via-state-augmentation}

In \cref{subsec:governing-equation-and-path-integral} and \cref{subsec:path-probability-measure-via-the-path-integral-representation}, we develop the path density under a Markov assumption. In practice, the drift $f_\lambda^\theta$ may take a short history window as input \citep{Tashiro+21,Price+25}. This does not violate the Markov formulation: the process becomes Markov after augmenting the state with past time steps. For clarity, we consider a drift that depends on the current and one previous time step; the extension to longer windows is straightforward.

Consider a drift of the form $f_\lambda^\theta(u_\lambda(t^{(n)}), u_\lambda(t^{(n-1)}))$. As in \cref{subsec:path-probability-measure-via-the-path-integral-representation}, we suppress the spatial argument $x$ for notational simplicity. Define the augmented state $U^{(n)} \in \mathbb{R}^{2D}$ as
\begin{align}
    U^{(n)} := \begin{pmatrix} u_\lambda(t^{(n)}) \\ u_\lambda(t^{(n-1)}) \end{pmatrix}. \label{eq:augmented-state-definition}
\end{align}
The time evolution of $U^{(n)}$ is governed by
\begin{align}
    U^{(n+1)} = U^{(n)} + F(U^{(n)}) \Delta t + G\, \Xi^{(n)} \Delta t, \label{eq:augmented-dynamics}
\end{align}
where $\Xi^{(n)} := (\xi(t^{(n)}), 0)^\top \in \mathbb{R}^{2D}$ with $\xi(t^{(n)}) \sim \mathcal{N}(0, \frac{1}{\Delta t} I_D)$ denoting Gaussian white noise (\cref{subsec:path-probability-measure-via-the-path-integral-representation}). The drift $F: \mathbb{R}^{2D} \to \mathbb{R}^{2D}$ and noise coefficient matrix $G \in \mathbb{R}^{2D \times 2D}$ are
\begin{align}
    F(U^{(n)}) := \begin{pmatrix} f_\lambda^\theta(u_\lambda(t^{(n)}), u_\lambda(t^{(n-1)})) \\ \dfrac{u_\lambda(t^{(n)}) - u_\lambda(t^{(n-1)})}{\Delta t} \end{pmatrix}, \quad
    G := \begin{pmatrix} \sigma_\lambda I_D & 0 \\ 0 & 0 \end{pmatrix}. \label{eq:augmented-drift-and-noise}
\end{align}
The first component of $F$ is the learned drift acting on the history window, while the second component shifts the previous state forward deterministically. The matrix $G$ injects stochasticity only into the current state; past states evolve without additional noise. This structure ensures that the augmented dynamics in Eq.~(\ref{eq:augmented-dynamics}) is Markov in $U^{(n)}$.

For reverse-$\lambda$ sampling in Eq.~(\ref{eq:reverse-sde-spatiotemporal}), we need the path density $p_\lambda(\{u_\lambda\}_t)$. We define it using the original variables $\{u_\lambda(t^{(n)})\}_{n=0}^{N_t}$ induced by the Markov chain in the augmented space. Following Eq.~(\ref{eq:transition-probability-density}), the transition density is
\begin{align}
    p_\lambda\!\left(u_\lambda(t^{(n+1)}) \mid u_\lambda(t^{(n)}), u_\lambda(t^{(n-1)})\right) = \mathcal{N}\!\Big(
    u_\lambda(t^{(n+1)})\,;\, u_\lambda(t^{(n)}) + f_\lambda^\theta(u_\lambda(t^{(n)}), u_\lambda(t^{(n-1)}))\,\Delta t,\;\sigma_\lambda^2 \Delta t\, I_D\Big). \label{eq:history-transition-density}
\end{align}
Let $r_\lambda(u_\lambda(t^{(0)}), u_\lambda(t^{(-1)}))$ denote the initial density of the history window. For temporal boundary handling, we set the missing past state $u_\lambda(t^{(-1)})$ via zero padding (\cref{subsec:neural-network-architecture}). The path density then factorizes as
\begin{align}
    p_\lambda(\{u_\lambda\}_t) = r_\lambda(u_\lambda(t^{(0)}), u_\lambda(t^{(-1)})) \prod_{n=0}^{N_t-1} p_\lambda\!\left(u_\lambda(t^{(n+1)}) \mid u_\lambda(t^{(n)}), u_\lambda(t^{(n-1)})\right). \label{eq:history-factorized-path-density}
\end{align}
The path score $\nabla_{u_\lambda}\ln p_\lambda(\{u_\lambda\}_t)$ can be computed by automatic differentiation of the log-density in Eq.~(\ref{eq:history-factorized-path-density}).

\subsection{Surrogate Model Obtained by KL Divergence Minimization}
\label{subsec:surrogate-model-obtained-by-kl-divergence-minimization}

We derive the surrogate model $f_\lambda^\theta$ as the minimizer of the KL divergence $D_{\mathrm{KL}}(q_\lambda \| p_\lambda)$ in Eq.~(\ref{eq:kl-divergence}) using the discrete-time path density (\cref{subsec:path-probability-measure-via-the-path-integral-representation}). This yields a regression problem and characterizes the optimal drift as a conditional expectation of temporal increments.

At $\lambda=0$, we assume the existence of a true physical (deterministic) model $f_{0}^{\rm true}$ whose temporal evolution is governed by
\begin{align}
    \partial_t u_0(t) = f_{0}^{\rm true}(u_0(t)).
\end{align}
As in \cref{sec:proposed-method} and \cref{subsec:path-probability-measure-via-the-path-integral-representation}, we discretize physical time by $t^{(n)} = n\Delta t$ for $n=0,\dots,N_t$. For notational simplicity, spatial dependence on $x$ is taken into account but not written explicitly in the equations below. The state $u_\lambda(t^{(n)})\in\mathbb{R}^{D}$ denotes the spatially discretized state (including channels).

To derive the KL divergence $D_{\mathrm{KL}}(q_\lambda \| p_\lambda)$, we need the probability densities $q_\lambda(\{u_\lambda\}_t)$ and $p_\lambda(\{u_\lambda\}_t)$. The latter is derived in Eqs.~(\ref{eq:transition-probability-density}) and (\ref{eq:path-integral-formal-representation}). According to \cref{sec:proposed-method}, $q_\lambda(\{u_\lambda\}_t)$ is obtained through the conditional density at each discrete time step (cf.~Eq.~(\ref{eq:conditional-density})):
\begin{align}
    q_\lambda\!\left(u_\lambda(t^{(n)}) \mid u_0(t^{(n)})\right) &= \mathcal{N}\!\left(u_\lambda(t^{(n)})\,;\, \mathcal{C}_\lambda u_0(t^{(n)}), \Sigma_\lambda\right), \label{eq:q-transient-conditional} \\
    q_\lambda(\{u_\lambda\}_t \mid \{u_0\}_t) &= \prod_{n=0}^{N_t} q_\lambda\!\left(u_\lambda(t^{(n)}) \mid u_0(t^{(n)})\right). \label{eq:q-path-conditional-product}
\end{align}
Let $q_{\rm d}(\{u_0\}_t)$ denote the (unknown) distribution of true paths at $\lambda=0$. We define the joint density by
\begin{align}
    q_\lambda^{\rm joint}(\{u_\lambda\}_t,\{u_0\}_t) := q_\lambda(\{u_\lambda\}_t \mid \{u_0\}_t)\,q_{\rm d}(\{u_0\}_t). \label{eq:q-joint}
\end{align}
Marginalizing $q_\lambda^{\rm joint}$ with respect to $\{u_0\}_t$, we obtain $q_\lambda(\{u_\lambda\}_t)$.

The KL divergence is
\begin{align}
    D_{\mathrm{KL}}(q_\lambda \| p_\lambda) &= \int \mathcal{D}u_\lambda\, q_\lambda(\{u_\lambda\}_t)\, \ln\frac{q_\lambda(\{u_\lambda\}_t)}{p_\lambda(\{u_\lambda\}_t)}, \\
    &= \mathbb{E}_{(\{u_\lambda\}_t,\{u_0\}_t)\sim q_\lambda^{\rm joint}} \left[-\ln p_\lambda(\{u_\lambda\}_t)\right] + \mathrm{const.}, \label{eq:kl-reduction}
\end{align}
where $q_\lambda(\{u_\lambda\}_t)$ does not depend on $\theta$, and $\mathrm{const.}$ collects all terms independent of $\theta$. Using Eqs.~(\ref{eq:transition-probability-density}) and (\ref{eq:path-integral-formal-representation}), the $\theta$-dependent part of $D_{\mathrm{KL}}(q_\lambda \| p_\lambda)$ is
\begin{align}
    \mathbb{E}_{(\{u_\lambda\}_t,\{u_0\}_t)\sim q_\lambda^{\rm joint}} \left[ \frac{1}{2\sigma_\lambda^2\Delta t} \sum_{n=0}^{N_t-1} \left\|u_\lambda(t^{(n+1)}) - u_\lambda(t^{(n)}) - f_\lambda^\theta(u_\lambda(t^{(n)}))\Delta t\right\|^2 \right]. \label{eq:kl-to-increments-mse}
\end{align}
Here, $\|\cdot\|$ denotes the Euclidean norm over all spatial degrees of freedom and channels at a fixed time step; this is equivalent to integrating the channel-wise norm over space in the main-text notation (up to the $\Delta x$ discretization factor). This is the explicit form of the loss function in Eq.~(\ref{eq:training-loss}). Minimizing this loss yields the pointwise optimal drift satisfying
\begin{align}
    f_\lambda^*(u)\Delta t = \mathbb{E}\!\left[u_\lambda(t^{(n+1)})-u_\lambda(t^{(n)})\;\middle|\;u_\lambda(t^{(n)})=u\right], \label{eq:optimal-drift-conditional-mean-increment}
\end{align}
where the conditional expectation is taken under $q_\lambda^{\rm joint}$.

We next connect this optimal drift to the fine-resolution (i.e., not coarse-grained) dynamics. Since $q_\lambda(u_\lambda(t^{(n+1)})\mid u_0(t^{(n+1)}))=\mathcal{N}(\mathcal{C}_\lambda u_0(t^{(n+1)}),\Sigma_\lambda)$ in Eq.~(\ref{eq:q-transient-conditional}), the law of iterated expectations leads to
\begin{align}
    \mathbb{E}\!\left[u_\lambda(t^{(n+1)}) \mid u_\lambda(t^{(n)})\right] &= \mathbb{E}\!\left[\mathbb{E}\!\left[u_\lambda(t^{(n+1)}) \mid u_0(t^{(n+1)})\right]\middle| u_\lambda(t^{(n)})\right] = \mathbb{E}\!\left[\mathcal{C}_\lambda u_0(t^{(n+1)}) \mid u_\lambda(t^{(n)})\right].
\end{align}
Substituting this into Eq.~(\ref{eq:optimal-drift-conditional-mean-increment}) yields
\begin{align}
    f_\lambda^*(u_\lambda(t^{(n)}))\Delta t &= \mathbb{E}\!\left[\mathcal{C}_\lambda u_0(t^{(n+1)}) \mid u_\lambda(t^{(n)})\right] - u_\lambda(t^{(n)}), \\
    &= \mathbb{E}\!\left[\mathcal{C}_\lambda [u_0(t^{(n+1)})-u_0(t^{(n)})] \mid u_\lambda(t^{(n)})\right] + \left(\mathbb{E}\!\left[\mathcal{C}_\lambda u_0(t^{(n)}) \mid u_\lambda(t^{(n)})\right] - u_\lambda(t^{(n)})\right), \label{eq:conditional-expectation-increment}
\end{align}
where we use the linearity of $\mathcal{C}_\lambda$ (\cref{subsec:coarse-graining-operator-and-covariance-matrix}). The first term is the conditional expectation of the coarse-grained true increment, while the second term is a denoising correction induced by conditioning on the \emph{noisy} coarse-grained state $u_\lambda(t^{(n)})$:
\begin{align}
    \delta^{(n)}_\lambda := \mathbb{E}\!\left[\mathcal{C}_\lambda u_0(t^{(n)}) \mid u_\lambda(t^{(n)})\right] - u_\lambda(t^{(n)}). \label{eq:denoising-correction}
\end{align}
Using the true dynamics at $\lambda=0$, we can also write
\begin{align}
    \mathcal{C}_\lambda [u_0(t^{(n+1)})-u_0(t^{(n)})] = \mathcal{C}_\lambda\!\left(\int_{t^{(n)}}^{t^{(n)}+\Delta t} f_0^{\rm true}(u_0(t'))\,\mathrm{d}t'\right), \label{eq:coarse-grained-increment}
\end{align}
so the first term in Eq.~(\ref{eq:conditional-expectation-increment}) becomes a conditional expectation of the coarse-grained true increment. If $\Delta t$ is sufficiently small, it is natural to approximate
\begin{align}
    \mathcal{C}_\lambda [u_0(t^{(n+1)})-u_0(t^{(n)})] \approx \mathcal{C}_\lambda f_0^{\rm true}(u_0(t^{(n)}))\,\Delta t. \label{eq:coarse-grained-instantaneous-increment}
\end{align}
Equation~(\ref{eq:conditional-expectation-increment}) shows that, under the assumed $q_\lambda^{\rm joint}$, the optimal drift generally includes the denoising correction $\delta^{(n)}_\lambda$. This correction is further discussed below and in \cref{subsec:specification-of-the-noise-amplitude-sigma-lambda}.

The correction $\delta^{(n)}_\lambda$ in Eq.~(\ref{eq:denoising-correction}) depends on the (unknown) posterior distribution of $u_0(t^{(n)}) \mid u_\lambda(t^{(n)})$ induced by $q_{\rm d}$, and hence is not available in closed form without further assumptions. Nevertheless, because Eq.~(\ref{eq:q-transient-conditional}) is an additive Gaussian noise model, the correction admits a general identity \citep[i.e., Tweedie-type formula;][]{Efron11}. Let $q_\lambda^{(n)}(u)$ denote the marginal density of $u_\lambda(t^{(n)})$ under $q_\lambda^{\rm joint}$. Assuming sufficient regularity, the denoising correction can be written as
\begin{align}
    \mathbb{E}\!\left[\mathcal{C}_\lambda u_0(t^{(n)})\mid u_\lambda(t^{(n)})=u_\lambda\right] - u_\lambda = \Sigma_\lambda \nabla_{u_\lambda}\ln q_\lambda^{(n)}(u_\lambda), \label{eq:tweedie-correction}
\end{align}
where $\nabla_{u_\lambda}$ denotes the gradient with respect to $u_\lambda$ at the fixed time step $t^{(n)}$.

As a further specialization, if one approximates $u_0(t^{(n)})$ as Gaussian, $u_0(t^{(n)})\sim\mathcal{N}(\overline{u_{\rm d}},\Sigma_{\rm d})$ (the subscript $\rm d$ denotes data), the posterior mean becomes
\begin{align}
    \mathbb{E}\!\left[\mathcal{C}_\lambda u_0(t^{(n)})\mid u_\lambda(t^{(n)}) = u_\lambda\right] &= \mathcal{C}_\lambda \overline{u_{\rm d}} + \Sigma_{\rm d}' \left( \Sigma_{\rm d}' + \Sigma_\lambda \right)^{-1} \left(u_\lambda-\mathcal{C}_\lambda \overline{u_{\rm d}}\right), \\
    \Sigma_{\rm d}' &:= \mathcal{C}_\lambda \Sigma_{\rm d} \mathcal{C}_\lambda^{\top}.
\end{align}
Therefore, the correction $\delta^{(n)}_\lambda$ in Eq.~(\ref{eq:denoising-correction}) becomes
\begin{align}
    \left.\delta^{(n)}_\lambda \right|_{u_\lambda(t^{(n)})=u_\lambda} = -\Sigma_\lambda\left(\Sigma_{\rm d}'+\Sigma_\lambda\right)^{-1} \left(u_\lambda-\mathcal{C}_\lambda \overline{u_{\rm d}}\right). \label{eq:denoising-correction-gaussian}
\end{align}

Equation~(\ref{eq:conditional-expectation-increment}) refines the main-text interpretation of Eq.~(\ref{eq:conditional-expectation-drift}) by explicitly separating (i) the conditional expectation of the coarse-grained true increment and (ii) the denoising correction $\delta^{(n)}_\lambda$ that arises because $u_\lambda(t^{(n)})$ is a noisy observation of $\mathcal{C}_\lambda u_0(t^{(n)})$ under $q_\lambda^{\rm joint}$. The simpler form in Eq.~(\ref{eq:conditional-expectation-drift}) captures the dominant contribution and provides useful intuition, while the full expression in Eq.~(\ref{eq:conditional-expectation-increment}) determines the exact optimal drift. As $\lambda \to 0$ (and hence $\Sigma_\lambda \to 0$), $\delta^{(n)}_\lambda$ vanishes and the two expressions coincide. An interpretation of $\delta^{(n)}_\lambda$ is further discussed in \cref{subsec:specification-of-the-noise-amplitude-sigma-lambda}.

The first term in Eq.~(\ref{eq:conditional-expectation-increment}) (or its approximation~(\ref{eq:conditional-expectation-drift})) evaluates the time evolution on the fine-resolution field---including all Fourier modes---before applying the coarse-graining operator $\mathcal{C}_\lambda$. This ordering is crucial: small-scale dynamics contribute to large-scale evolution through the conditional expectation. In contrast, evaluating the drift on an already coarse-grained field (i.e., $f_0^{\rm true}(\mathcal{C}_\lambda u_0)$) would lose such contributions. The trained drift $f_\lambda^\theta$ therefore emulates large-scale variability while implicitly accounting for the influence of eliminated small-scale components. This conclusion is consistent with the general result that non-perturbative RG transformations preserve long-range correlation structures \citep{Carosso20,Cotler+Rezchikov23}.

\subsection{Specification of the Noise Amplitude $\sigma_\lambda$}
\label{subsec:specification-of-the-noise-amplitude-sigma-lambda}

We discuss the estimation of the noise amplitude $\sigma_\lambda$. This quantity represents the uncertainty arising from the elimination of small-scale components through coarse-graining \citep{Carosso20,Cotler+Rezchikov23}. Although $\sigma_\lambda$ could in principle be treated as a learnable parameter, we determine it from the solution of the forward process in Eq.~(\ref{eq:forward-rg-sde}) that describes coarse-graining.

From the solution in Eq.~(\ref{eq:ou-solution-spatiotemporal}), the coarse-grained state at each $(\,\cdot\,, t^{(n)})$ is
\begin{align}
  u_\lambda(\,\cdot\,, t^{(n)}) = \mathcal{C}_\lambda u_0(\,\cdot\,, t^{(n)}) + \sqrt{\Sigma_\lambda} \epsilon(\,\cdot\,, t^{(n)}),\label{eq:coarse-grained-u-at-grid-x-t}
\end{align}
where $\epsilon(\,\cdot\,, t^{(n)}) \sim \mathcal{N}(0,I_D)$. Taking a temporal difference yields
\begin{align}
    u_\lambda(\,\cdot\,, t^{(n+1)}) - u_\lambda(\,\cdot\,, t^{(n)}) = \mathcal{C}_\lambda \left[u_0(\,\cdot\,, t^{(n+1)}) - u_0(\,\cdot\,, t^{(n)})\right] + \sqrt{\Sigma_\lambda} \left[\epsilon(\,\cdot\,, t^{(n+1)}) - \epsilon(\,\cdot\,, t^{(n)})\right], \label{eq:temporal-diff-coarse-grained-u}
\end{align}
where we use the linearity of $\mathcal{C}_\lambda$ and $\sqrt{\Sigma_\lambda}$. By comparing with the governing equation~(\ref{eq:discrete-time-governing-equation}), we see that the first term on the right-hand side corresponds to the drift term $f_\lambda^\theta \Delta t$, while the second term corresponds to the noise term $\sigma_\lambda \xi \Delta t$. This finding suggests that coarse-graining through a stochastic process naturally induces a noise term in the temporal dynamics.

We now clarify the connection to the denoising correction $\delta^{(n)}_\lambda$ in Eqs.~(\ref{eq:conditional-expectation-increment}) and (\ref{eq:denoising-correction}). Taking the conditional expectation of Eq.~(\ref{eq:temporal-diff-coarse-grained-u}) under $q_\lambda^{\rm joint}$ in Eq.~(\ref{eq:q-joint}) yields
\begin{align}
    \mathbb{E}\!\left[u_\lambda(t^{(n+1)}) - u_\lambda(t^{(n)}) \mid u_\lambda(t^{(n)})\right]
    &= \mathbb{E}\!\left[\mathcal{C}_\lambda[u_0(t^{(n+1)}) - u_0(t^{(n)})] \mid u_\lambda(t^{(n)})\right] - \sqrt{\Sigma_\lambda}\,\mathbb{E}\!\left[\epsilon(t^{(n)}) \mid u_\lambda(t^{(n)})\right], \label{eq:optimal-drift-noise-form}
\end{align}
where we suppress the $x$ dependence for notational brevity. Since $\epsilon(t^{(n+1)})$ is independent of $u_\lambda(t^{(n)})$, its conditional expectation vanishes. However, the conditional expectation of $\epsilon(t^{(n)})$ remains because $\epsilon(t^{(n)})$ is correlated with $u_\lambda(t^{(n)})$ through the coarse-graining operation: $u_\lambda(t^{(n)}) = \mathcal{C}_\lambda u_0(t^{(n)}) + \sqrt{\Sigma_\lambda}\epsilon(t^{(n)})$. The resulting second term on the right-hand side equals the denoising correction $\delta^{(n)}_\lambda$ in Eq.~(\ref{eq:denoising-correction}). Indeed, following Eq.~(\ref{eq:optimal-drift-conditional-mean-increment}), the optimal drift satisfies $f_\lambda^* \Delta t = \mathbb{E}[u_\lambda(t^{(n+1)}) - u_\lambda(t^{(n)}) \mid u_\lambda(t^{(n)})]$, so Eq.~(\ref{eq:optimal-drift-noise-form}) is another equivalent form of $f_\lambda^*$. When the learned drift $f_\lambda^\theta$ is well-fitted, it automatically captures $\sqrt{\Sigma_\lambda}\,\mathbb{E}\!\left[\epsilon(t^{(n)}) \mid u_\lambda(t^{(n)})\right]$, suggesting that the next-time noise $\epsilon(t^{(n+1)})$ is the dominant irreducible stochasticity. This justifies matching the average variance (i.e., the trace) of $\sigma_\lambda^2 \Delta t I_D$ to $\Sigma_\lambda$.

To obtain a concrete value for $\sigma_\lambda$, we match the conditional covariances of the surrogate and data distributions. The surrogate model in Eq.~(\ref{eq:discrete-time-governing-equation}) has conditional covariance $\sigma_\lambda^2 \Delta t I_D$, where $I_D$ is the $D$-dimensional identity matrix. Under $q_\lambda^{\rm joint}$ in Eq.~(\ref{eq:q-joint}), we derive the conditional covariance of $u_\lambda(t^{(n+1)})$ given $u_\lambda(t^{(n)})$ from Eq.~(\ref{eq:temporal-diff-coarse-grained-u}). This covariance includes the uncertainty in predicting $\mathcal{C}_\lambda u_0(t^{(n+1)})$ and the irreducible contribution $\Sigma_\lambda$ from the next-time noise $\epsilon(t^{(n+1)})$. As argued above, the latter is expected to capture the dominant irreducible stochasticity when the drift is well-fitted. Matching the average variance (i.e., equating traces of covariance matrices) yields
\begin{align}
  \sigma_\lambda^2 = \frac{1}{\Delta t \cdot D} \mathrm{tr}(\Sigma_\lambda).
  \label{eq:sigma-trace-matching}
\end{align}
This indicates that $\sigma_\lambda$ characterizes the typical scale of coarse-graining noise. We use this setting of $\sigma_\lambda$ for all the experiments. Equation~(\ref{eq:sigma-trace-matching}) incorporates only the irreducible contribution $\Sigma_\lambda$ (and ignores the uncertainty in predicting $\mathcal{C}_\lambda u_0(t^{(n+1)})$), and thus provides a conservative estimate of $\sigma_\lambda$. We confirmed that moderate rescaling of $\sigma_\lambda$ (e.g., by a factor of 1.5) does not significantly affect the results.

We note the behavior of $\sigma_\lambda$ as $\lambda \to 0$. Following Eq.~(\ref{eq:coarse-graining-covariance-in-fourier}), $\Sigma_\lambda \to 0$ as $\lambda \to 0$, which implies $\sigma_\lambda \to 0$. Mathematically, no singularity arises in the path probability itself: the path density concentrates on deterministic paths as a well-defined distributional limit (\cref{subsec:path-probability-measure-via-the-path-integral-representation}). However, the path score $\nabla_{u_\lambda} \ln p_\lambda$ diverges in this limit. For numerical stability, we lower-bound $\Sigma_\lambda$ by $\Sigma_{\Delta\lambda}$, the covariance at the smallest non-zero diffusion scale ($\lambda\in[0,1]$ is discretized into 1,000 steps, so $\Delta\lambda = 10^{-3}$; see \cref{subsec:hyperparameters}). This regularization ensures $\sigma_\lambda > 0$ in Eq.~(\ref{eq:sigma-trace-matching}), preventing divergence of the path score.

\section{Details of Experimental Methods}
\label{sec:details-of-experimental-methods}

\subsection{Construction of Training Data}
\label{subsec:construction-of-training-data}

\paragraph{Lorenz-96 Model}
The Lorenz-96 model is a one-dimensional idealized model that represents zonal atmospheric dynamics \citep{Lorenz96}. Periodic boundary conditions are assumed, analogous to the longitudinal direction of the Earth, with $x \in [0, 2\pi)$. The state variables $X$ and $Y$ represent slow and fast variables in physical time $t$, respectively, and evolve according to
\begin{align}
    \frac{\mathrm{d}}{\mathrm{d}t} X_k &= -X_{k-1}\left(X_{k-2}-X_{k+1}\right)-X_k+F-\left(\frac{h c}{b}\right) \sum_{j=1}^{J} Y_{j,k}, \\
    \frac{\mathrm{d}}{\mathrm{d}t} Y_{j,k} &= -c b Y_{j+1,k}\left(Y_{j+2, k}-Y_{j-1,k}\right)-c Y_{j,k}+\frac{h c}{b} X_k.
\end{align}
The indices $j$ and $k$ denote grid-point locations on the one-dimensional spatial domain. For each $k$, $J$ fast variables $Y_{j,k}$ interact with the corresponding slow variable $X_k$ for $j = 1, \ldots, J$. To match the spatial resolution, each $X_k$ is replicated $J$ times so that both $X$ and $Y$ are defined on $KJ$ grid points. The ordering of grid points follows \citet{Russell+17}, where $X_k=X_{k+K}$, $Y_{j,k}=Y_{j, k+K}$ and $Y_{j, k}=Y_{j-J,k+1}$. Model parameters are chosen according to previous studies \citep{Lorenz96,Russell+17}, with $K = 32$, $J = 4$ (yielding a total of $KJ = 128$ grid points), $h = 1$, and $F = b = c = 10$. The parameter $b$ characterizes the amplitude ratio between $X$ and $Y$, while $c$ characterizes the timescale separation. With $b = c = 10$, the slow variable $X$ has amplitudes and timescales approximately one order of magnitude larger than those of the fast variable $Y$.

Training data for the Lorenz-96 model are generated by numerical integration using the Euler method with a time step of $5 \times 10^{-4}$. Initial conditions are randomly varied, and samples are collected after the system reaches a statistically stationary state, with a burn-in period of 30 time units. Each sample consists of two channels $(X, Y)$, with 64 snapshots in time (i.e., $N_t=63$ increments) and a spatial length of 128 grid points. The temporal sampling interval is set to 0.05 time units, corresponding to 100 integration steps. We verify that varying this sampling interval does not change the temporal power spectrum. A total of 3,000 samples are used for training and 100 samples with distinct initial conditions are used for testing. We confirm that the test results are not strongly dependent on the sample size.

\paragraph{Kolmogorov Flow}
The two-dimensional Kolmogorov flow is an incompressible flow system driven by sinusoidal forcing \citep[e.g.,][]{Obukhov+83} and is widely used in turbulence research and machine learning applications \citep[e.g.,][]{Lucas+Kerswell14,Kochkov+21,Shu+23}. The governing equations are
\begin{align}
    \frac{\partial \zeta}{\partial t} + \mathbf{V} \cdot \nabla_x \zeta = - \mu \zeta + \nu \nabla_x^2 \zeta - k_{\rm forcing} \cos (k_{\rm forcing} y), \\
    \zeta = \nabla_x^2 \psi \quad\text{and}\quad\mathbf{V} = \left(-\frac{\partial \psi}{\partial y}, \frac{\partial \psi}{\partial x}\right)^{\top}.
\end{align}
Here, $\nabla_x$ and $\nabla_x^2$ denote the gradient and Laplacian with respect to the spatial coordinates $(x,y)$, $\zeta$ the vorticity field, $\mathbf{V}$ the velocity field, and $\psi$ the stream function. Doubly periodic boundary conditions are assumed on $(x,y) \in [0, 2\pi)^2$. Model parameters are chosen following previous studies \citep{Kochkov+21,Shu+23}, with $\mu = 0.1$, $\nu = 1 \times 10^{-3}$, and $k_{\rm forcing} = 4$.

Spatial derivatives and nonlinear terms are computed using a pseudospectral method \citep[e.g.,][]{Canuto+88}, and time integration is performed using the explicit midpoint method, also known as the modified Euler method. The spatial grid resolution is set to $256 \times 256$, and the integration time step is $1 \times 10^{-3}$. Initial conditions are randomly varied, and samples are collected after the system reaches a statistically stationary state, with a burn-in period of 50 time units. Each sample consists of a single channel corresponding to $\zeta$, with 40 snapshots in time (i.e., $N_t=39$ increments) and a spatial resolution of $40 \times 40$. The spatial fields are resized using bicubic interpolation, and the temporal dimension is subsampled at an interval of 0.5 time units. We verify that changing the temporal sampling interval does not affect the temporal power spectrum. A total of 6,000 samples are used for training and 100 samples with distinct initial conditions are used for testing. We confirm that the test results are not strongly dependent on the sample size.

\paragraph{Data Preprocessing}
All state variables (i.e., $X, Y, \zeta$) are standardized to have zero mean and unit standard deviation. This normalization removes the zero-wavenumber constant mode and eliminates the need for scale-independent damping terms \citep{Cotler+Rezchikov23}.

\subsection{Neural Network Architecture}
\label{subsec:neural-network-architecture}

\paragraph{U-Net.}
We use a U-Net-based neural network architecture \citep{Ronneberger+15,Ho+20,Ho+22b,Saharia+22b,Saharia+22a}. The number of channels is first increased to 32 by convolution, after which the spatial resolution is progressively reduced. Self-attention mechanisms are then applied \citep{Vaswani+17}, and finally the spatial resolution is restored. For the Lorenz-96 model, one-dimensional convolutions are used to reduce the spatial resolution to 1/16 ($=1/2^4$) of the original size, while the number of channels is gradually increased to 64, 128, 128, and 192. For the Kolmogorov flow, two-dimensional convolutions are used to reduce the spatial resolution to 1/8 ($=1/2^3$) of the original size, with the number of channels increased to 64, 128, and 192. Downsampling is performed using convolutions with stride 2, and upsampling is performed using transposed convolutions. The Swish activation function is used throughout the network \citep{Ramachandran+17}. The diffusion scale $\lambda$ is provided via Feature-wise Linear Modulation (FiLM) conditioning \citep{Perez+18}, which modulates intermediate features through learned scale and shift parameters.

The input to the neural network consists of the state vectors at the current time and the four most recent past time steps. For the Lorenz-96 model, this corresponds to 10 input channels, representing $X$ and $Y$ over five time steps, while for the Kolmogorov flow the input consists of 5 channels, representing $\zeta$ over five time steps. All variables are concatenated along the channel dimension before being provided to the network. The network output is a finite-difference approximation of the time derivative in physical time $t$. In \cref{subsec:dependence-on-the-number-of-input-time-steps}, we vary the number of input time steps; the network architecture is kept fixed throughout.

\paragraph{FNO.}
As an alternative architecture, we also employ a Fourier Neural Operator (FNO) \citep{Li+21}. The FNO applies spectral convolutions in Fourier space with mode truncation, followed by pointwise nonlinearities. We use 8 Fourier layers for the Lorenz-96 model and 6 layers for the Kolmogorov flow, with a hidden width of 64 in both cases. The number of retained Fourier modes is 56 (out of 64 Nyquist modes) for the Lorenz-96 model and 16 (out of 20 Nyquist modes) for the Kolmogorov flow. The Swish activation and FiLM conditioning for $\lambda$ are the same as in the U-Net. The input and output formats are also identical. Unlike the U-Net, the FNO uses neither convolutions nor self-attention, providing a test of whether our proposed method works across different network architectures. Results with FNO are presented in \cref{subsec:experimental-results-with-fno}.

\paragraph{Temporal boundary handling.}
For the first few temporal indices where the history window is not fully available, we use zero padding for the missing past states.

\subsection{Hyperparameters}
\label{subsec:hyperparameters}

\paragraph{Validation set.}
For hyperparameter tuning, we use 100 samples with distinct initial conditions as the validation set for both the Lorenz-96 model and Kolmogorov flow, separate from both the training and test sets.

\paragraph{Coarse-graining.}
For the Lorenz-96 model, we set $\alpha = 0.1$ and $\beta = \sqrt{2}$, while for the Kolmogorov flow we set $\alpha = 0.3$ and $\beta = \sqrt{6}$. The Lorenz-96 model has 128 spatial grid points, which is approximately three times that of the Kolmogorov flow (i.e., 40 grid points per spatial dimension), and therefore a smaller value of $\alpha$ is used for the Lorenz-96 model. We do not set $\alpha$ exactly to $1/9$ ($=1/3^2$), because the Kolmogorov flow is a two-dimensional system and a direct comparison is not straightforward. These parameter values are determined by grid search, and we confirm that the results are not overly sensitive to the exact values. The diffusion scale is taken as $\lambda \in [0,1]$ ($\lambda_{\min}=0$ and $\lambda_{\max}=1$), and a discretization with 1,000 steps is used (i.e., $\Delta \lambda = 10^{-3}$).

\paragraph{Optimization.}
We train all models using Adam \citep{Kingma+15} with learning rate $2\times 10^{-4}$ and batch size 50 for the Lorenz-96 model and 40 for the Kolmogorov flow. We train for $3\times 10^4$ optimization iterations (no epoch-based schedule and no early stopping). We use the default Adam hyperparameters, with no weight decay and no gradient clipping. We use automatic mixed precision (AMP) with FP16. The DDPM baselines in \cref{subsec:baseline-models} use the same optimization settings.

\subsection{Baseline Models}\label{subsec:baseline-models}

As a score-based baseline, we train a denoising diffusion probabilistic model (DDPM) using the denoising score matching objective \citep{Ho+20,Vincent11}. We use the same neural network architecture as described in \cref{subsec:neural-network-architecture} and the same training hyperparameters as in \cref{subsec:hyperparameters}. The forward diffusion process is variance-preserving \citep{Ho+20,Song+21} with $\alpha=0$ and $\gamma=3$, i.e., $\partial_\lambda u_\lambda(x,t) = -\gamma u_\lambda(x,t) + \sqrt{2\gamma}\,\eta_\lambda(x,t)$, and we discretize $\lambda \in [0,1]$ into 1,000 steps. As in our proposed method, we do not use a $\lambda$-dependent noise schedule. Input time series are concatenated along the channel dimension, and we use the same zero padding at temporal boundaries as in \cref{subsec:neural-network-architecture}.

\paragraph{Baseline for Simulation (Conditional Prediction)}
For simulation, we follow the conditional diffusion setup used in GenCast \citep{Price+25}. The denoiser estimates the score of the standardized temporal difference (residual) conditioned on the five most recent clean states (the current state and the four preceding time steps). Since this baseline uses $\alpha=0$, it does not support simulations at $\lambda>0$; we therefore compare simulations only at $\lambda=0$. At inference time, we generate the residual one step at a time and roll out in physical time by sliding the conditioning window.

\paragraph{Baseline for Unconditional Generation}
For unconditional generation, we model the joint distribution over full spatiotemporal fields and draw samples via DDPM-style reverse-$\lambda$ integration with 1,000 steps \citep{Ho+22b}, using the same reverse-$\lambda$ discretization as the proposed method (\cref{subsec:detailed-implementation}). To evaluate the score at each temporal grid $t^{(n)}$, the denoiser takes a local temporal window $(u_\lambda(t^{(n)}), u_\lambda(t^{(n-1)}), \ldots, u_\lambda(t^{(n-4)}))$ from the noisy sample, with zero padding where $n < 4$. Unlike the simulation baseline, no clean conditioning context is used---all five inputs are noisy. During reverse-$\lambda$ integration, the score is evaluated at every $t^{(n)}$ and the full spatiotemporal field is updated simultaneously.

\subsection{Evaluation Metrics}
\label{subsec:evaluation-metrics}

We use the relative $L^2$ error for short-term prediction accuracy and the relative spectral error for long-term statistical consistency \citep{Jiang+23}, defined as:
\begin{align}
    \text{Relative } L^2 \text{ Error}(t) &:= \frac{\| u(\cdot,t) - \hat{u}(\cdot,t)\|_2}{\| u(\cdot,t) \|_2},  \label{eq:l2-err-definition} \\
    \text{Relative Spectral Error} &:= \frac{\| \; |\mathcal{G}[u]|^2 - |\mathcal{G}[\hat{u}]|^2 \; \|_1}{\| \; |\mathcal{G}[u]|^2 \; \|_1}, \label{eq:spec-err-definition}
\end{align}
where $u$ and $\hat{u}$ denote state variables from physics-based and surrogate simulations, respectively, and $\mathcal{G}$ denotes the Fourier transform over $x$ and $t$. Unless otherwise stated, we compare $u$ and $\hat{u}$ at the same diffusion scale $\lambda$. The only exception in the main text is the low-resolution (LR) column in Table~\ref{table:main_text_error_super_resolution}, which compares a coarse-grained simulation result at $\lambda=0.2$ against the $\lambda=0$ physics-based reference. We omit the term ``relative'' below and in the main text.

Both metrics are computed after standardizing $u$ and $\hat{u}$ with shared normalization parameters. The $L^2$ error is evaluated at each $t$ and averaged over test samples. For the spectral error, we average the spatiotemporal power spectral densities (PSDs) over test samples and compute the relative difference. To estimate statistical uncertainty, we train the neural network five times with different random initializations and compute standard deviations across these runs.

For the $L^2$ error, surrogate simulations use the same initial conditions as the reference. For the spectral error, we discard the first 10 segments (each spanning the temporal length of one sample) and compute PSDs from the remaining segments; discarding 100 segments instead of 10 yields similar spectral errors and confirms long-rollout stability.

For $\lambda > 0$, reference data $u_\lambda$ are constructed by applying $\mathcal{C}_\lambda$ to $u_0$. Since coarse-grained references exclude noise, we set $\sigma_\lambda = 0$ in Eq.~(\ref{eq:governing-eq-along-t}) for evaluation. Although simulations with $\sigma_\lambda \ne 0$ are stable, we omit noise to allow clear comparison with the reference. Simulation results with $\sigma_\lambda > 0$ are provided in \cref{subsec:simulation-with-noise}.

\subsection{Detailed Implementation}
\label{subsec:detailed-implementation}

This subsection provides implementation details for training (\cref{alg:training}), forward simulation (\cref{alg:prediction}), and generation/super-resolution (\cref{alg:generation}), including numerical techniques for stable integration. Each algorithm references the corresponding equation.

For the discretized stochastic equations in Algorithms~\ref{alg:prediction} and~\ref{alg:generation}, we adopt the white Gaussian noise convention \citep[e.g.,][]{Shiraishi23}: the noise variables $\xi$ and $\eta_\lambda$ are sampled from $\mathcal{N}(0, \frac{1}{\Delta t}I_D)$ and $\mathcal{N}(0, \frac{1}{\Delta\lambda}I_{D'})$, respectively (see \cref{subsec:notation} for details), so that the increments $\xi\,\Delta t$ and $\eta_\lambda\,\Delta\lambda$ have covariance matrices $\Delta t I_D$ and $\Delta\lambda I_{D'}$.

\begin{algorithm}[tb]
    \caption{Training of Predictor-Driven Diffusion Model}
    \label{alg:training}
    \textbf{Input:} training dataset of sample paths; functions $\mathcal{C}_\lambda$, $\Sigma_\lambda$, $\sigma_\lambda$; neural network $f_\lambda^\theta$ \\
    \textbf{Output:} optimal parameters $\theta^*$
    \begin{algorithmic}[1]
        \REPEAT
        \STATE Sample a minibatch of spatiotemporal paths $\{\{u_0\}_t^{(b)}\}_{b=1}^{B}$ from the training dataset
        \STATE Sample $\lambda^{(b)} \sim \mathcal{U}(0,1)$ and $\epsilon^{(b)} \sim \mathcal{N}(0,I_{D'})$ for each $b$
        \STATE Compute $\mathcal{C}_{\lambda^{(b)}}$, $\Sigma_{\lambda^{(b)}}$, and $\sigma_{\lambda^{(b)}}$ for each $b$
        \STATE $u_{\lambda^{(b)}}^{(b)} \leftarrow \mathcal{C}_{\lambda^{(b)}} u_0^{(b)} + \sqrt{\Sigma_{\lambda^{(b)}}}\,\epsilon^{(b)}$ {\footnotesize in Eq.~(\ref{eq:ou-solution-spatiotemporal})}
        \STATE $\mathcal{L}(\theta) \leftarrow \frac{1}{B}\sum_{b=1}^{B}\frac{1}{2\sigma_{\lambda^{(b)}}^2}\int_{\rm x,t}\left\| \partial_t u_{\lambda^{(b)}}^{(b)}-f_{\lambda^{(b)}}^\theta(u_{\lambda^{(b)}}^{(b)})\right\|^2$ {\footnotesize in Eq.~(\ref{eq:training-loss})}
        \STATE Update $\theta$ using $\nabla_\theta \mathcal{L}(\theta)$
        \UNTIL{convergence}
    \end{algorithmic}
\end{algorithm}

\begin{algorithm}[tb]
    \caption{Forward Simulation in $t$ at Fixed $\lambda$}
    \label{alg:prediction}
    \textbf{Input:} fine-resolution initial condition $u_0(\,\cdot\,,t^{(0)})$; diffusion scale $\lambda$; functions $\mathcal{C}_\lambda$, $\sigma_\lambda$; learned neural network $f_\lambda^\theta$; time step $\Delta t$; number of increments $N_t$ \\
    \textbf{Output:} simulated path $\{u_\lambda(\,\cdot\,,t)\}_{t=t^{(0)}}^{t^{(N_t)}}$, where $t^{(0)} = 0$ and $t^{(N_t)}=N_t \Delta t$
    \begin{algorithmic}[1]
        \STATE Set $\sigma_\lambda \leftarrow 0$ for deterministic simulation, otherwise compute $\sigma_\lambda$ for the given $\lambda$
        \STATE Initialize $u_\lambda(\,\cdot\,,t^{(0)}) \leftarrow \mathcal{C}_\lambda u_0(\,\cdot\,,t^{(0)})$
        \FOR{$n=0$ {\bfseries to} $N_t - 1$}
        \STATE Sample $\xi \sim \mathcal{N}(0, \frac{1}{\Delta t}I_D)$ if $\sigma_\lambda > 0$; otherwise set $\xi \leftarrow 0$ 
        \STATE $u_\lambda\left(\,\cdot\,,t^{(n+1)}\right) \leftarrow u_\lambda(\,\cdot\,,t^{(n)}) + f_\lambda^\theta(u_\lambda)\,\Delta t + \sigma_\lambda\,\xi \Delta t$ {\footnotesize in Eq.~(\ref{eq:governing-eq-along-t})}
        \ENDFOR
    \end{algorithmic}
\end{algorithm}

For generation and super-resolution (\cref{alg:generation}), we adopt predictor-corrector sampling \citep{Song+21}. Each iteration consists of a predictor step that integrates the reverse-$\lambda$ equation~(\ref{eq:reverse-sde-spatiotemporal}), followed by Langevin corrector steps that refine the sample toward the target distribution. Following \citep[Algorithm 4]{Song+21}, each corrector step draws $z \sim \mathcal{N}(0,I_{D'})$, recomputes the score $s_\lambda := \nabla_{u_\lambda} \ln p_\lambda(\{u_\lambda\}_t)$, defines $\chi := 2\left(\kappa\,\|z\|_2/\|s_\lambda\|_2\right)^2$ ($\kappa$: signal-to-noise ratio), and updates
\begin{equation}
    u_\lambda' = u_\lambda + \chi s_\lambda + \sqrt{2\chi}\, z.
    \label{eq:langevin-corrector}
\end{equation}
We use 3 corrector steps per predictor step; in our experiments, results were not sensitive to this choice (1 or 2 steps yielded similar performance). We select the signal-to-noise ratio parameter $\kappa$ on the validation set by grid search over $[0,2]$ with step size 0.05 to minimize the spectral error in Eq.~(\ref{eq:spec-err-definition}): $\kappa=0.7$ for the Lorenz-96 model and $\kappa=0.3$ for the Kolmogorov flow. Both the proposed method and the baseline use these values in the test phase.

The predictor step employs an exponential time differencing (ETD) scheme \citep{Hochbruck+Ostermann10,Lord+Tambue19}, a standard approach for stiff equations. In reverse-$\lambda$ integration, the Laplacian coefficient $\alpha\left\|k\right\|^2$ varies by orders of magnitude across Fourier modes, requiring small step sizes for high-wavenumber modes under explicit methods. Let $\widetilde{u}_\lambda(k,t)=\mathcal{F}[u_\lambda(\cdot,t)](k)$, $\widetilde{s}_\lambda(k,t)=\mathcal{F}[s_\lambda(\cdot,t)](k)$, and $\widetilde{\eta}_\lambda(k,t)=\mathcal{F}[\eta_\lambda(\cdot,t)](k)$ denote spatial Fourier transforms at each fixed $t$ (\cref{subsec:coarse-graining-operator-and-covariance-matrix}), where our convention is $\mathcal{F}[\nabla_x^2 u_\lambda](k,t)=-\left\|k\right\|^2\widetilde{u}_\lambda(k,t)$. The update proceeds as
\begin{align}
    \widetilde{u}_{\lambda-\Delta\lambda}(k,t) = e^{\alpha\left\|k\right\|^2\Delta\lambda}\widetilde{u}_\lambda(k,t) + \varphi(\alpha\left\|k\right\|^2\Delta\lambda)\,\beta^2 \widetilde{s}_\lambda(k,t)\Delta\lambda + \beta\,\widetilde{\eta}_\lambda(k,t)\Delta\lambda, \label{eq:exponential-integrator}
\end{align}
where $\varphi(z) = (e^z - 1)/z$ with $\varphi(0)=1$ by continuity. Since integration proceeds backward in $\lambda$, the Laplacian term $\alpha \nabla_x^2 u_\lambda$ acts as anti-diffusion; the exponential factor $e^{\alpha\left\|k\right\|^2\Delta\lambda} > 1$ assigns a larger amplification rate to high-wavenumber modes, counteracting the scale-dependent damping in the forward coarse-graining. The update is performed in Fourier space and the updated field $\widetilde{u}_{\lambda-\Delta\lambda}$ is then transformed back to real space. Inspired by \citet{Lord+Tambue19}, the deterministic and stochastic terms are treated separately, with the $\varphi$-function applied only to the deterministic component (i.e., the score term). For the noise term, we use the Euler--Maruyama scheme to preserve the spatially white noise structure assumed in Eq.~(\ref{eq:reverse-sde-spatiotemporal}); applying $e^{\alpha\left\|k\right\|^2\Delta\lambda}$ to the noise term would introduce $k$-dependent amplification, inconsistent with this assumption. For the baseline DDPM, we apply the same ETD scheme with $\alpha\left\|k\right\|^2$ replaced by a constant $\gamma$ (\cref{subsec:baseline-models}), which corresponds to scale-independent amplification.

The path density in Eq.~(\ref{eq:path-probability-density}) contains the initial density $r_\lambda$. This $r_\lambda$ is not needed during training, where it drops out of the KL divergence as a constant in Eq.~(\ref{eq:kl-divergence}), nor during forward-$t$ simulation, which requires only the drift. However, generation via Eq.~(\ref{eq:reverse-sde-spatiotemporal}) requires computing the path score $s_\lambda = \nabla_{u_\lambda} \ln p_\lambda$. At $t = t^{(0)}$, this score includes the boundary term $\nabla_{u_\lambda} \ln r_\lambda$. We do not model this term explicitly; instead, after each predictor step we overwrite the boundary value at $t^{(0)}$ by linear extrapolation, $u_{\lambda-\Delta\lambda}(x,t^{(0)}) \leftarrow 2\,u_{\lambda-\Delta\lambda}(x,t^{(1)}) - u_{\lambda-\Delta\lambda}(x,t^{(2)})$, and apply the same extrapolation within the Langevin corrector steps. This extrapolation stabilizes reverse-$\lambda$ sampling near the temporal boundary. Even when the drift depends on the past four time steps (\cref{subsec:neural-network-architecture}), overwriting only $t^{(0)}$ was sufficient in our experiments. Explicitly modeling $r_\lambda$ (or its score) remains a direction for future work, particularly for statistically unsteady systems (see Limitations in \cref{sec:conclusions}).

\begin{algorithm}[tb]
    \caption{Generation / Super-Resolution (Reverse-$\lambda$ Integration)}
    \label{alg:generation}
    \textbf{Input:} $\lambda_{\max}>\lambda_{\min}$; $\alpha,\beta,\kappa$; function $\Sigma_\lambda$; neural network $f_\lambda^\theta$; step size $\Delta\lambda>0$; coarse-grained path $\left\{u_\lambda^{\rm LR}\right\}_{t}$ (for super-resolution) \\
    \textbf{Output:} generated / super-resolved path $\left\{u_{\lambda_{\min}}\right\}_{t}$
    \begin{algorithmic}[1]
        \STATE \textbf{(Generation)} Sample $u_{\lambda_{\max}}(\,\cdot\,,t^{(n)}) {\sim} \mathcal{N}(0,\Sigma_{\lambda_{\max}})$ for $n=0,\ldots,N_t$
        \STATE \textbf{(Super-resolution)} Initialize $\left\{u_{\lambda_{\max}}\right\}_{t} \leftarrow \left\{u_\lambda^{\rm LR}\right\}_{t}$
        \FOR{$\lambda=\lambda_{\max},\,\lambda_{\max}-\Delta\lambda,\,\dots,\,\lambda_{\min}+\Delta\lambda$}
        \STATE Compute score $s_\lambda \leftarrow \nabla_{u_\lambda} \ln p_\lambda(\{u_\lambda\}_t)$
        \STATE Sample $\eta_\lambda \sim \mathcal{N}(0, \frac{1}{\Delta\lambda}I_{D'})$
        \STATE \textbf{Predictor:} Update $u_{\lambda-\Delta\lambda}$ via ETD in Eq.~(\ref{eq:exponential-integrator}); extrapolate at $t^{(0)}$
        \STATE \textbf{Corrector:} (with $\kappa$ as specified)
        \FOR{$m=1$ {\bfseries to} $3$}
        \STATE Compute score $s_\lambda \leftarrow \nabla_{u_\lambda} \ln p_\lambda(\{u_\lambda\}_t)$
        \STATE Sample $z \sim \mathcal{N}(0, I_{D'})$
        \STATE Update $u_\lambda$ via Langevin step in Eq.~(\ref{eq:langevin-corrector}); extrapolate at $t^{(0)}$
        \ENDFOR
        \ENDFOR
    \end{algorithmic}
\end{algorithm}

\subsection{Computational Cost}
\label{subsec:computational-cost}

Each training iteration samples $\lambda$ and computes a loss on time derivatives at that $\lambda$ (\cref{alg:training}); the per-iteration cost is comparable (up to constants) to DDPM-style training when using the same backbone and similar batching strategies \citep{Ho+20}. Simulation at a fixed $\lambda$ is inexpensive: the trained drift $f_\lambda^\theta$ is evaluated once per physical-time step to advance the state, with no backpropagation required (\cref{alg:prediction}). However, generation and super-resolution are more costly because reverse-$\lambda$ sampling requires computing the path score via automatic differentiation (\cref{alg:generation}). We discuss this cost in more detail below.

In our implementation, a spatiotemporal sample $\{u_\lambda\}_t$ is regarded as a single tensor, analogous to a multi-channel image, and the score is obtained as the input gradient of the scalar log-density with respect to this tensor. This is a vector--Jacobian product computed by reverse-mode automatic differentiation \citep{Baydin+18}. Importantly, we do \emph{not} backpropagate through a physical-time integrator that repeatedly applies $f_\lambda^\theta$ to produce a rollout.

Reverse-mode differentiation of a scalar objective typically incurs a small constant-factor overhead over the corresponding forward evaluation, while its storage requirement can grow with the number of intermediate activations \citep{Baydin+18}. In our experiments, we did not observe memory bottlenecks, but memory can become the dominant constraint as the temporal length and model size increase. If needed, standard checkpointing strategies can trade additional computation for reduced activation memory \citep{Chen+16}.

Overall, sampling cost scales approximately linearly with the number of reverse-$\lambda$ steps and the temporal length $N_t$ (since the tensor size grows with $N_t$). We have not optimized the sampler for wall-clock performance, so there is room for improvement. For instance, reducing the number of reverse-$\lambda$ steps via sampler or noise scheduling is a direct way to lower cost \citep{Karras+22,Sheshmani+25,Masuki+Ashida25}; we leave such optimization for future work.

\newpage

\section{Details of Experimental Results}
\label{app:experimental-results}

This appendix discusses the details of the experimental results from \cref{sec:experiments}, mainly baseline comparisons.

\subsection{Simulation}
\label{subsec:details-of-experimental-results-simulation}

\begin{figure*}[h]
    \vskip 0.3in
    \begin{center}
        \includegraphics[width=\textwidth]{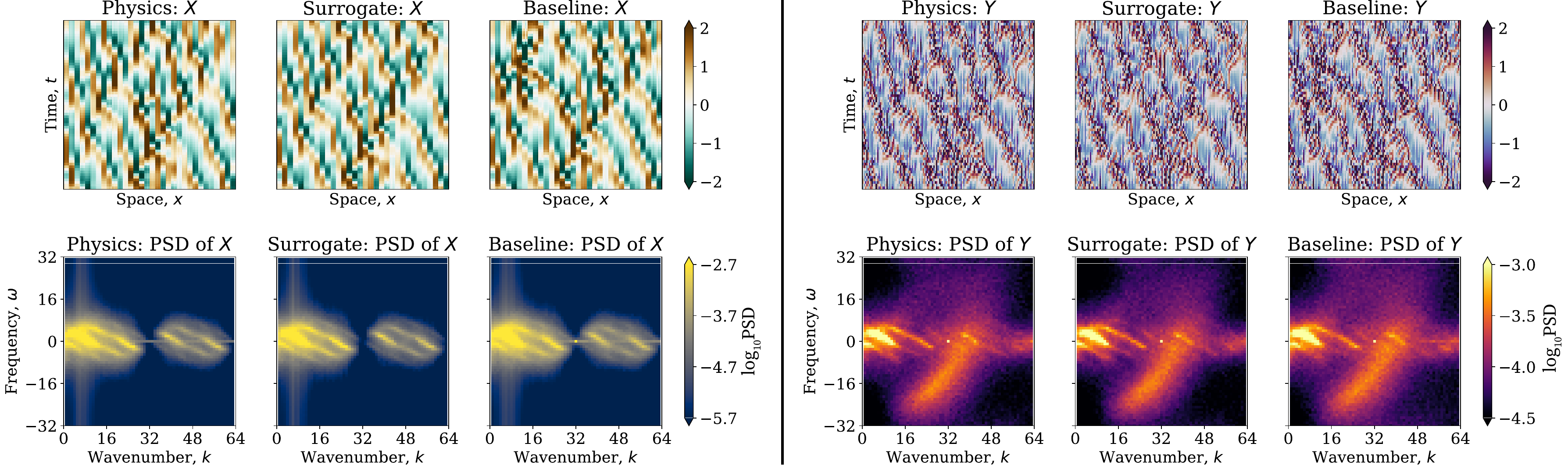}
        \caption{Simulation results at $\lambda = 0$ for the Lorenz-96 model, comparing our surrogate model with physics-based simulation and the DDPM baseline. Left: slow variable $X$; right: fast variable $Y$. Top: each panel shows physics-based simulation, surrogate prediction, and DDPM baseline; bottom: corresponding spatiotemporal power spectral densities (PSDs; computed from 100 samples). Both learned models accurately reproduce the spatiotemporal patterns and spectral statistics. All quantities are non-dimensionalized.}
        \label{fig:appendix_lorenz96_simulation_vs_baseline}
    \end{center}
\end{figure*}

\begin{figure*}[th]
    \vskip 0.0in
    \begin{center}
        \includegraphics[width=0.75\textwidth]{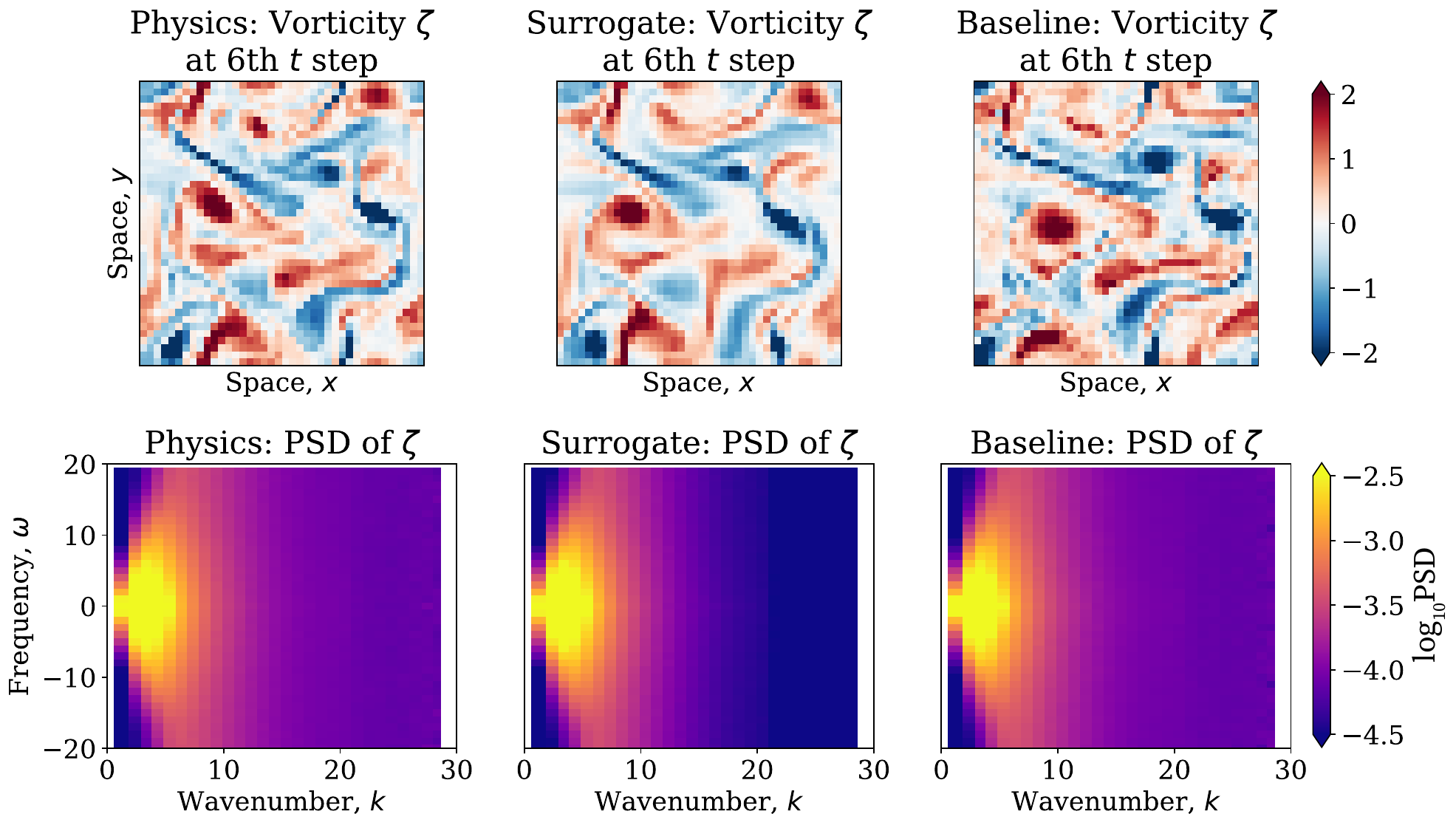}
        \caption{Simulation results at $\lambda = 0$ for the Kolmogorov flow, comparing our surrogate model with physics-based simulation and the DDPM baseline. Top row: vorticity $\zeta$ snapshots at the sixth time step; bottom row: spatiotemporal power spectral densities (PSD; computed from 100 samples). Left to right: physics-based simulation, surrogate prediction, and DDPM baseline. Both learned models reproduce the characteristic vorticity patterns and spectra. All quantities are non-dimensionalized.}
        \label{fig:appendix_kolmogorov_simulation_vs_baseline}
    \end{center}
\end{figure*}

Figures~\ref{fig:appendix_lorenz96_simulation_vs_baseline} and~\ref{fig:appendix_kolmogorov_simulation_vs_baseline} show simulation results for the Lorenz-96 model and Kolmogorov flow at diffusion scale $\lambda = 0$, respectively. As a baseline, we train a DDPM with denoising score matching conditioned on the same number of input time steps as our model (\cref{subsec:baseline-models}). Both our model (labeled as Surrogate) and the baseline produce predictions comparable to physics-based simulations. These results support the quantitative evaluation reported in Table~\ref{table:main_text_error_simulation}.

\begin{figure*}[hb]
    \vskip 0.1in
    \begin{center}
        \includegraphics[width=0.7\textwidth]{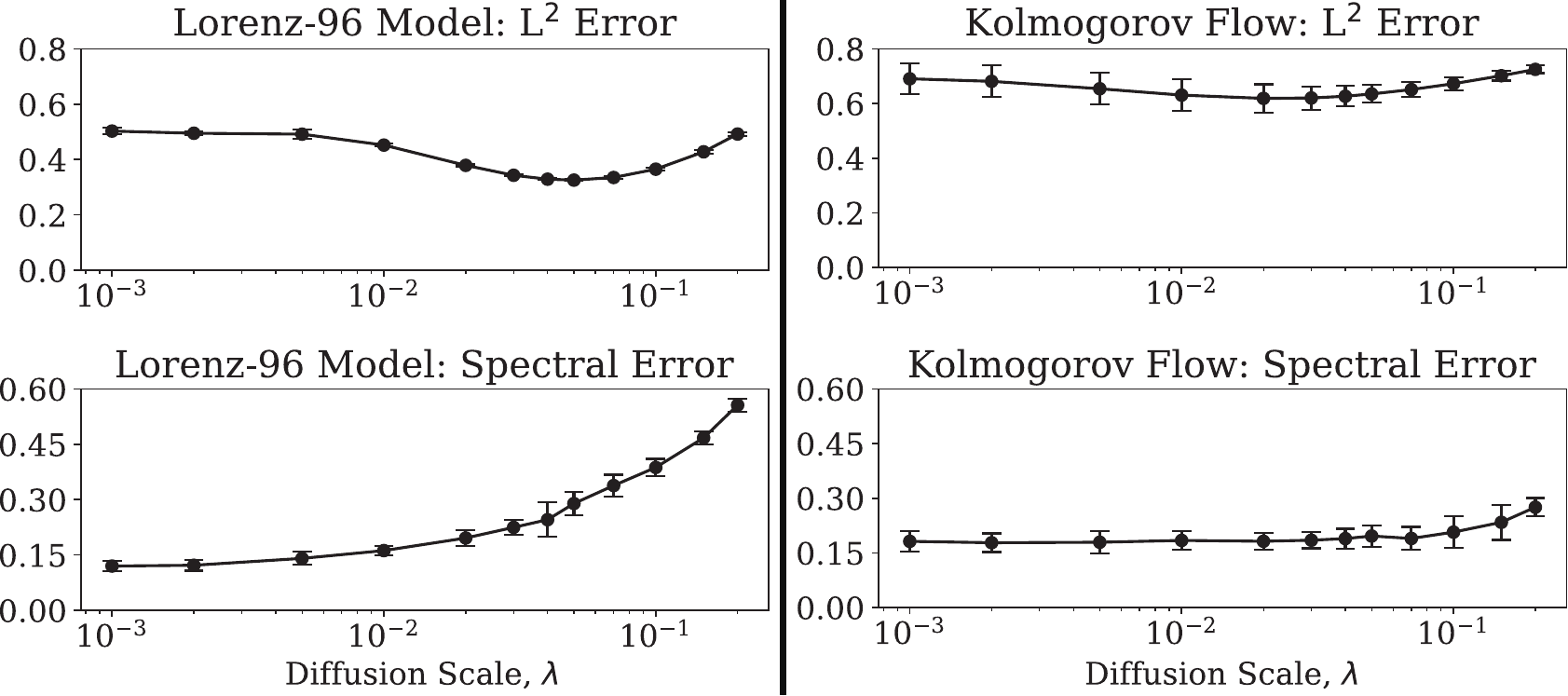}
        \caption{Dependence of evaluation metrics on $\lambda$ for the Lorenz-96 model and Kolmogorov flow. For each $\lambda$, we evaluate the surrogate prediction against a physics-based simulation coarse-grained to the same $\lambda$, as described in \cref{subsec:evaluation-metrics}. This differs from the LR column in Table~\ref{table:main_text_error_super_resolution}, which evaluates the LR result at $\lambda=0.2$ against the $\lambda=0$ physics-based simulation. To use a log scale, a small offset of $10^{-3}$ is added to $\lambda$ (e.g., $\lambda = 0$ is displayed as $\lambda = 10^{-3}$). The $L^2$ error is evaluated at the sixth time step. Error bars indicate means $\pm$ standard deviations across five training runs. All quantities are non-dimensionalized.}
        \label{fig:appendix_diffusion_scale_dependence}
    \end{center}
\end{figure*}

Unlike the baseline, our model can also perform prediction at different values of $\lambda$. Figure~\ref{fig:appendix_diffusion_scale_dependence} shows the $L^2$ error and spectral error for the Lorenz-96 model and Kolmogorov flow over the range $\lambda \in [0, 0.2]$. The spectral error increases with $\lambda$, reflecting the difficulty of reproducing long-term spatiotemporal patterns as coarse-graining progresses. We emphasize that all predictions over the range $\lambda \in [0, 0.2]$ are performed by a single trained model (\cref{sec:proposed-method}). Although the spectral error reaches approximately 0.6 at $\lambda=0.2$ for the Lorenz-96 model, this can be reduced to approximately 0.3 by increasing the number of input time steps (\cref{subsec:dependence-on-the-number-of-input-time-steps}).

The $L^2$ error (at the sixth time step), in contrast, depends relatively weakly on $\lambda$ (Fig.~\ref{fig:appendix_diffusion_scale_dependence}). In chaotic systems, short-term prediction errors saturate quickly regardless of the diffusion scale. For the Lorenz-96 model, the $L^2$ error initially decreases at intermediate $\lambda$ values. This occurs because coarse-graining preferentially removes small-scale components associated with the fast variable $Y$, resulting in an apparent error reduction. Although we do not inject noise here (\cref{subsec:experimental-setup,subsec:evaluation-metrics}), in a fully stochastic setting the injected noise would represent the effect of the eliminated small-scale dynamics.

\newpage

\newpage

\subsection{Spatiotemporal Data Generation}
\label{subsec:details-of-experimental-results-spatiotemporal-data-generation}

\begin{figure*}[tbh]
    \vskip 0.2in
    \begin{center}
        \includegraphics[width=\textwidth]{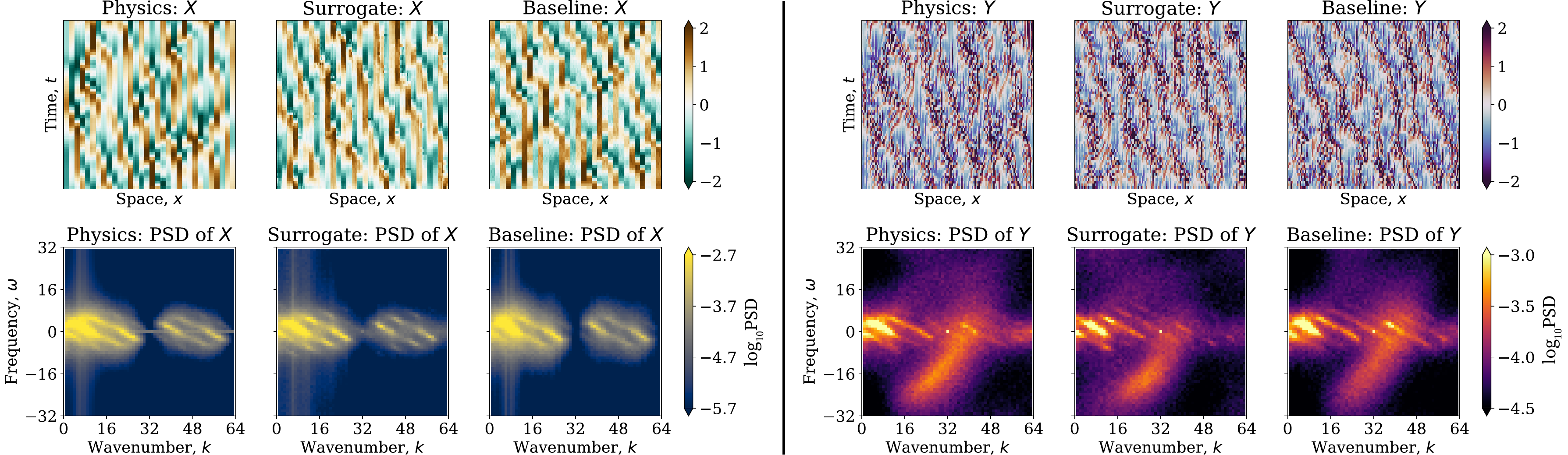}
        \caption{Unconditional generation results for the Lorenz-96 model, comparing our surrogate model with physics-based simulation and the DDPM baseline. Left: slow variable $X$; right: fast variable $Y$. Top: each panel shows physics-based simulation, surrogate generation, and DDPM baseline; bottom: corresponding power spectral densities (PSDs) computed from 100 samples. Both methods generate samples with spectral statistics matching the physics-based reference. All quantities are non-dimensionalized.}
        \label{fig:appendix_lorenz96_generation}
    \end{center}
\end{figure*}

\begin{figure*}[tbh]
    \vskip 0.2in
    \begin{center}
        \includegraphics[width=0.75\textwidth]{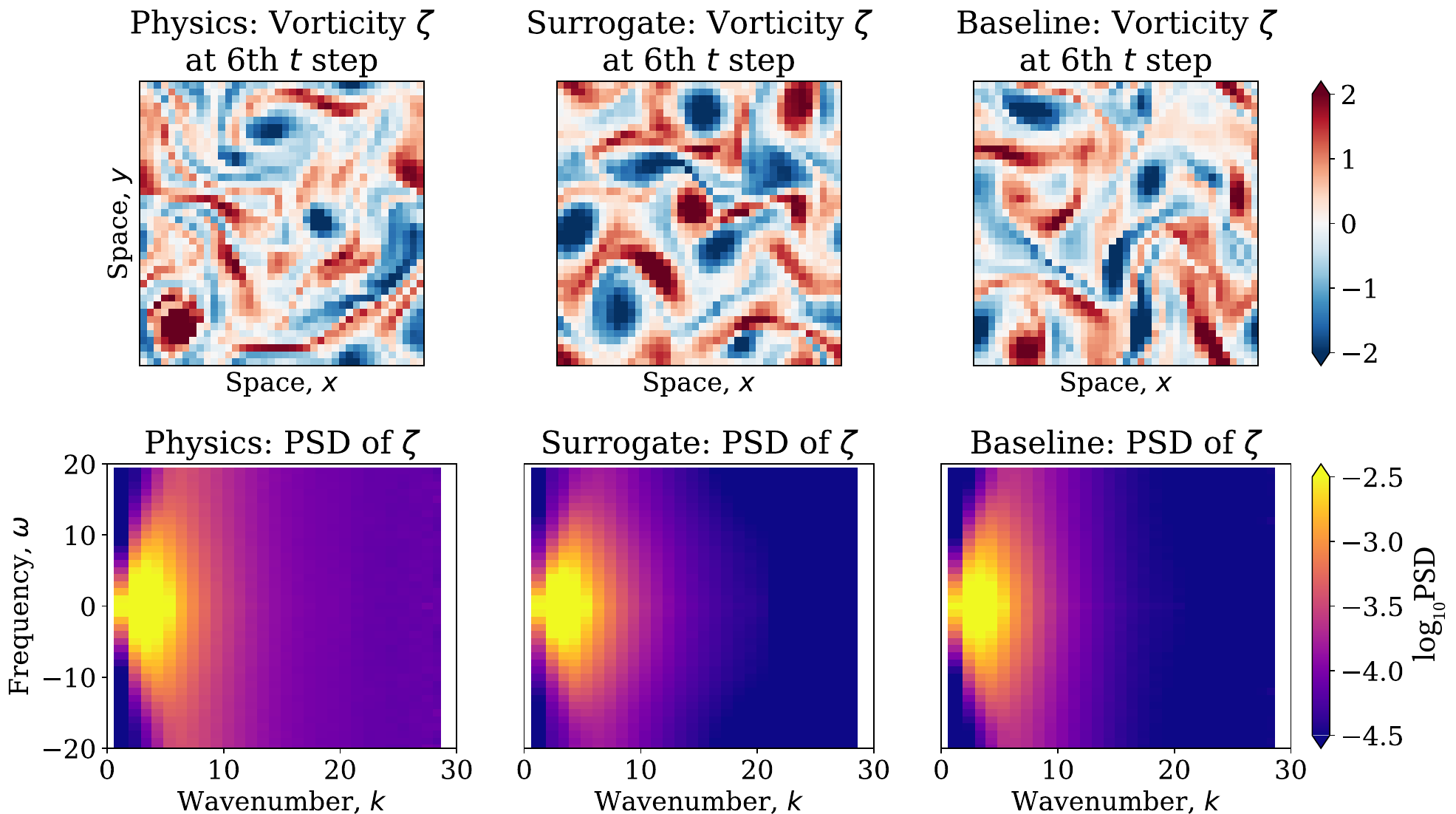}
        \caption{Unconditional generation results for the Kolmogorov flow, comparing our surrogate model with physics-based simulation and the DDPM baseline. Top row: vorticity $\zeta$ snapshots at the sixth time step; bottom row: spatiotemporal power spectral densities (PSD; computed from 100 samples). Left to right: physics-based simulation, surrogate generation, and DDPM baseline. Both methods reproduce the characteristic vorticity patterns and spectral statistics. All quantities are non-dimensionalized.}
        \label{fig:appendix_kolmogorov_generation}
    \end{center}
\end{figure*}

Figures~\ref{fig:appendix_lorenz96_generation} and~\ref{fig:appendix_kolmogorov_generation} show unconditional generation results for the Lorenz-96 model and Kolmogorov flow, respectively. As a baseline, we use a DDPM trained with denoising score matching (\cref{subsec:baseline-models}). Each figure displays a representative sample along with the spatiotemporal PSD computed from 100 samples. Both our model and the baseline generate samples that visually resemble those from physics-based simulations, and the spatiotemporal spectra are well reproduced. These results support the quantitative evaluation reported in Table~\ref{table:main_text_error_generation}.

\newpage

\section{Additional Experimental Results}
\label{app:additional-experimental-results}

This appendix reports additional experiments to further analyze the proposed framework. For conciseness, we primarily present results for the Lorenz-96 model; qualitatively similar results were obtained for the Kolmogorov flow.

\subsection{Simulation with Noise}
\label{subsec:simulation-with-noise}

\begin{figure*}[hb]
    \vskip 0.4in
    \begin{center}
        \includegraphics[width=\textwidth]{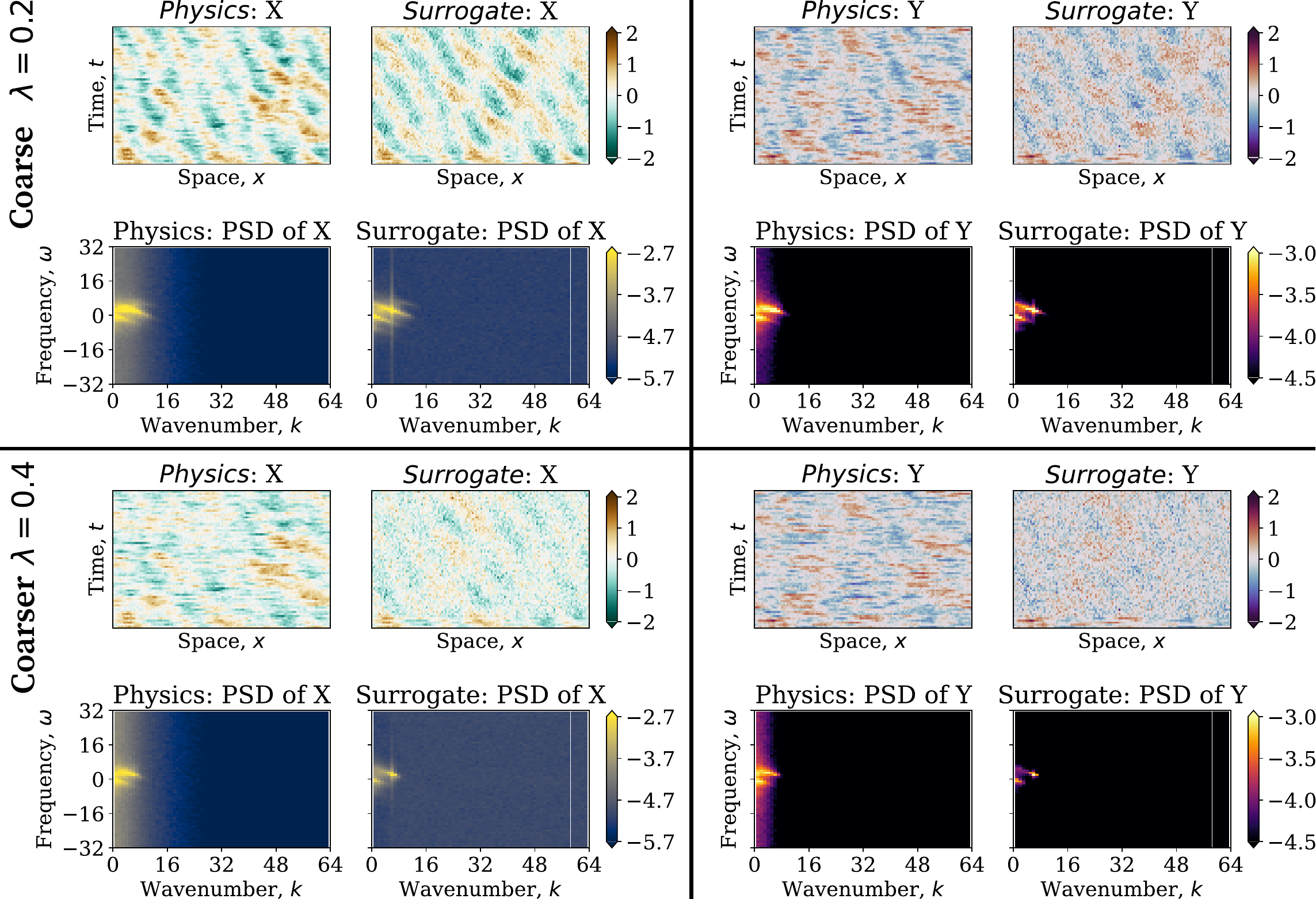}
        \caption{Simulation results at $\lambda=0.2$ and 0.4 for the Lorenz-96 model, where noise is injected during simulations (i.e., $\sigma_\lambda > 0$ in Eq.~(\ref{eq:governing-eq-along-t})). For spatiotemporal PSDs, $\log_{10}$ is applied to their magnitudes. All quantities are non-dimensionalized.}
        \label{fig:appendix_lorenz96_simulation_noise_injected}
    \end{center}
\end{figure*}

In this subsection only, we solve the temporal evolution equation~(\ref{eq:governing-eq-along-t}) with noise ($\sigma_\lambda > 0$) following Algorithm~\ref{alg:prediction}. The reference data are constructed by adding noise to physics-based simulation results as in Eq.~(\ref{eq:ou-solution-spatiotemporal}) and Algorithm~\ref{alg:training}. Since the noise amplitude is quite small when $\lambda \approx 0$, we illustrate the cases $\lambda=0.2$ and $\lambda=0.4$ in Fig.~\ref{fig:appendix_lorenz96_simulation_noise_injected}.

The model remains numerically stable even with noise injection. At $\lambda=0.2$, the model makes similar predictions to those of the reference. At $\lambda = 0.4$, however, the signal-to-noise ratio decreases, making comparison with the reference more difficult. Unlike scale-dependent coarse-graining (i.e., the Laplacian damping), this noise has statistically uniform amplitude across all Fourier modes, overwhelming large- and small-scale components and leaving the signal dominated by noise.

\newpage

\begin{figure*}[tbh]
    \begin{center}
        \includegraphics[width=\textwidth]{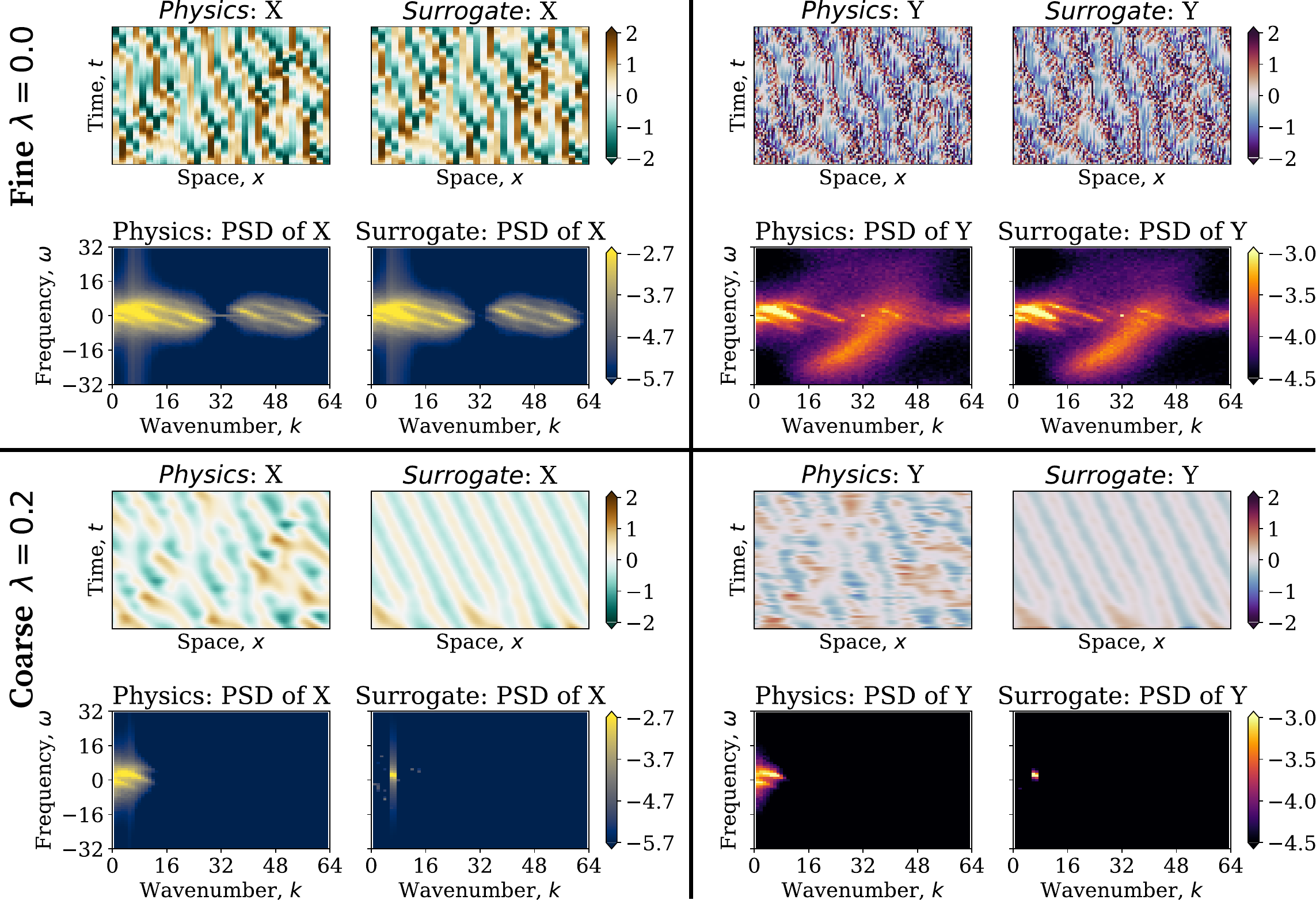}
        \caption{Simulation results for the Lorenz-96 model using only the current time step as input (window size = 1). Top two rows: $\lambda = 0$; bottom two rows: $\lambda=0.2$. Left to right: physics-based simulation, surrogate prediction, and spatiotemporal PSDs. At $\lambda = 0$, prediction is accurate since the system is Markov. At $\lambda=0.2$, a single input step fails to capture large-scale variability, producing monotonous patterns. PSDs are shown on a $\log_{10}$ scale. All quantities are non-dimensionalized.}
        \label{fig:appendix_lorenz96_simulation_window1}
    \end{center}
    \vskip 0.2in
\end{figure*}

\subsection{Dependence on the Number of Input Time Steps}
\label{subsec:dependence-on-the-number-of-input-time-steps}

It is common in time-series modeling to use multiple past time steps as inputs \citep[e.g.,][]{Tashiro+21,Yuan+24,Price+25}. However, our choice of five input steps may appear large compared with those in surrogate models for more complex atmospheric flows \citep{Price+25}. Here we discuss this choice based on the Mori--Zwanzig formalism \citep{Zwanzig61,Mori65,Kubo91}.

Figure~\ref{fig:appendix_lorenz96_simulation_window1} shows results using only the current time step as input. At $\lambda = 0$, this model yields accurate predictions since the Lorenz-96 model is Markov at fine resolution (\cref{subsec:construction-of-training-data}). At $\lambda=0.2$, however, a single input step fails to reproduce large-scale variability, producing monotonous patterns instead. In contrast, with five input steps, such variability is reproduced (Fig.~\ref{fig:main_text_lorenz96_simulation}).

\begin{figure*}[bh]
    \vskip 0.4in
    \begin{center}
        \includegraphics[width=0.75\textwidth]{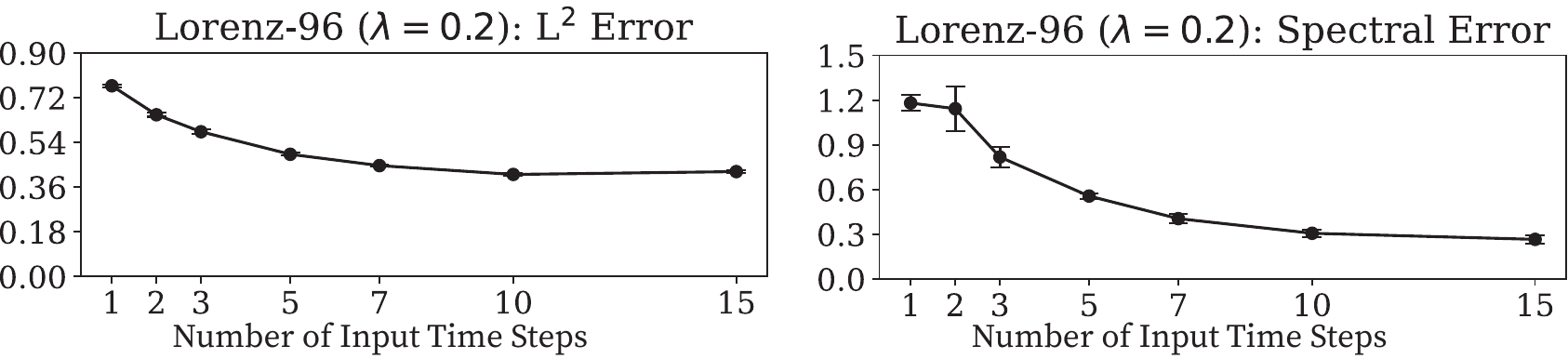}
        \caption{Dependence of prediction accuracy on the number of input time steps (window size) for the Lorenz-96 model at $\lambda=0.2$. Left: $L^2$ error; right: spectral error. The $L^2$ error is evaluated at the sixth time step. Error bars indicate means $\pm$ standard deviations across five training runs. All quantities are non-dimensionalized. Both metrics decrease as the window size increases, with spectral error dropping from approximately 0.6 (5 steps) to 0.3 (15 steps). This trend is consistent with the Mori--Zwanzig formalism: coarse-graining breaks the Markov assumption, requiring longer history to capture large-scale dynamics.}
        \label{fig:appendix_lorenz96_window_size_dependence}
    \end{center}
\end{figure*}

This result suggests that reproducing large-scale temporal evolution requires a sufficiently long history when small-scale components are eliminated. Indeed, the prediction error decreases as the number of input steps increases (Fig.~\ref{fig:appendix_lorenz96_window_size_dependence}): increasing from 5 to 15 steps reduces the spectral error from approximately 0.6 to 0.3.

These results are consistent with the Mori--Zwanzig projection operator formalism \citep{Zwanzig61,Mori65,Kubo91}. When small-scale components are eliminated by coarse-graining, the feedback they provide to large-scale dynamics must be represented using only the past history of large-scale variables, thereby breaking the Markov property. In time-series modeling, multiple past states are often provided as inputs \citep[e.g.,][]{Tashiro+21,Yuan+24,Price+25}, likely because including history improves reproduction of large-scale variability.

In the main text (\cref{sec:experiments}), we use five input time steps to adequately reproduce coarse-grained dynamics. Using fewer steps simplifies the preparation of initial conditions and reduces computational cost, leading to a trade-off between accuracy and efficiency; we adopt five steps as a compromise.

Furthermore, unconditional generation accuracy also improves with more input steps (not shown in detail). In unconditional generation, the reverse process in Eq.~(\ref{eq:reverse-sde-spatiotemporal}) is integrated toward smaller $\lambda$ (from $\lambda_{\max}$ to $\lambda_{\min}$). Since dynamics become more strongly history-dependent at large $\lambda$, accurate score estimation requires multiple input steps.

\newpage

\subsection{Laplacian Term in a Variance-Preserving Process}\label{subsec:effect-of-the-laplacian}

We compare two forward processes: a variance-preserving process \citep{Ho+20,Song+21} and its Laplacian-augmented extension. This tests whether the Laplacian yields scale-selective suppression even in the presence of uniform damping. Specifically, we consider the following forward equation,
\begin{align}
    \partial_\lambda u_\lambda(x,t) &= \left( \alpha\nabla_x^2 - \gamma \right) u_\lambda(x,t) +\beta \eta_\lambda(x,t). \label{eq:forward-sde-with-gamma}
\end{align}
We compare two cases: $\alpha = 0$ (no Laplacian) and $\alpha = 0.1$ (with Laplacian, as in the main experiments). In both cases, we set $\gamma = 3.0$ and $\beta = \sqrt{2\gamma}$ following the baseline setting (\cref{subsec:baseline-models}), which makes the process variance-preserving if $\alpha = 0$. All models are trained using the KL divergence objective (Algorithm~\ref{alg:training}), not score-based learning.

\begin{figure*}[h]
    \vskip 0.3in
    \begin{center}
        \includegraphics[width=0.85\textwidth]{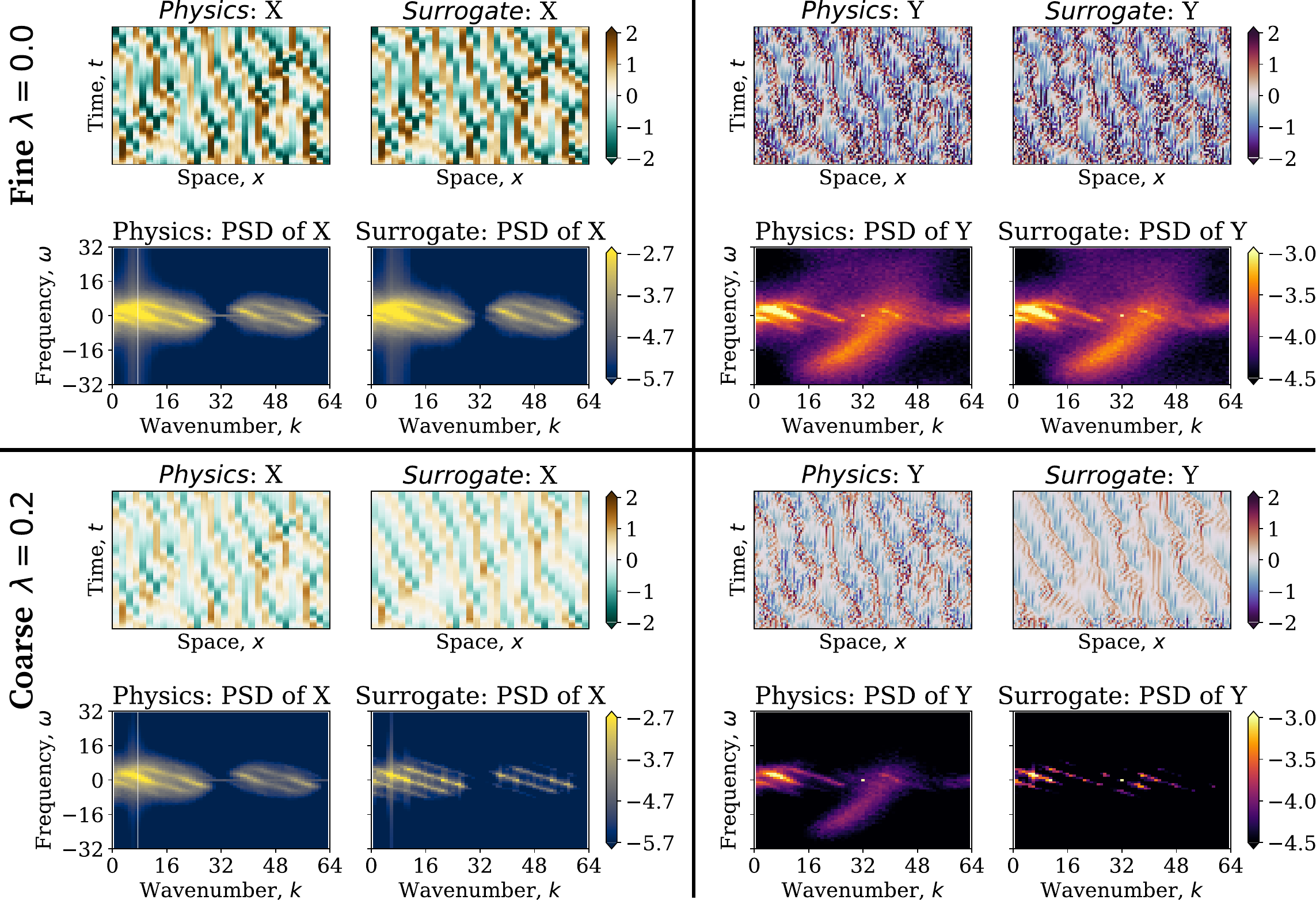}
        \caption{Simulation results for the Lorenz-96 model, where the forward process is described by Eq.~(\ref{eq:forward-sde-with-gamma}) with $\alpha=0$ (no Laplacian), $\gamma=3$, and $\beta=\sqrt{2\gamma}$. For spatiotemporal PSDs, $\log_{10}$ is applied to their magnitudes. All quantities are non-dimensionalized.}
        \label{fig:appendix_lorenz96_simulation_ablation_without_laplacian}
    \end{center}
\end{figure*}

\begin{figure*}[t]
    \vskip 0.2in
    \begin{center}
        \includegraphics[width=\textwidth]{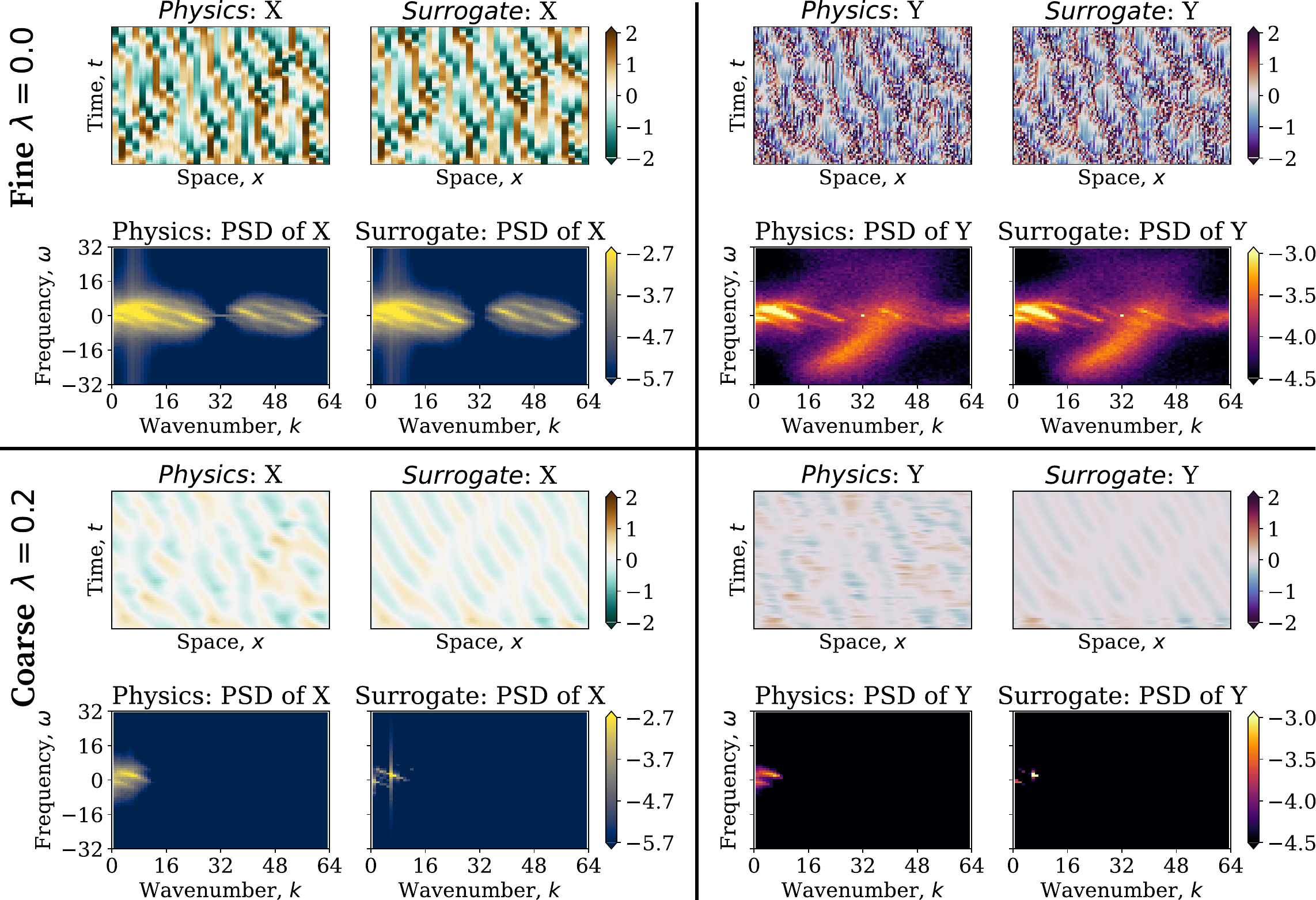}
        \caption{Simulation results for the Lorenz-96 model, where the forward process is described by Eq.~(\ref{eq:forward-sde-with-gamma}) with $\alpha=0.1$, $\gamma=3.0$, and $\beta=\sqrt{2\gamma}$. For spatiotemporal PSDs, $\log_{10}$ is applied to their magnitudes. All quantities are non-dimensionalized.}
        \label{fig:appendix_lorenz96_simulation_ablation_with_laplacian}
    \end{center}
    \vskip 0.2in
\end{figure*}

Without the Laplacian ($\alpha = 0$), all Fourier modes are damped at the same rate, so the relative magnitudes of large- and small-scale components remain unchanged across $\lambda$ (Fig.~\ref{fig:appendix_lorenz96_simulation_ablation_without_laplacian}). In contrast, with the Laplacian, small-scale components are sufficiently damped even when scale-independent damping $-\gamma u_\lambda(x,t)$ is included in Eq.~(\ref{eq:forward-sde-with-gamma}) (see Fig.~\ref{fig:appendix_lorenz96_simulation_ablation_with_laplacian}). These results suggest that our method works even in the presence of uniform damping.

\newpage

\subsection{Effect of Noise}
\label{subsec:effect-of-noise}

The forward process in Eq.~(\ref{eq:forward-rg-sde}) consists of a Laplacian term and a noise term. In \cref{subsec:effect-of-the-laplacian}, we studied the role of the Laplacian term; here, we ablate noise injection to examine its effect on simulation, generation, and super-resolution. Specifically, we train a model using deterministically coarse-grained data:
\begin{align}
    u_\lambda(x,t) = \mathcal{C}_\lambda u_0(x,t),
\end{align}
which is constructed by setting $\epsilon \equiv 0$ in Eq.~(\ref{eq:ou-solution-spatiotemporal}), i.e., $u_\lambda = \mathcal{C}_\lambda u_0 + \sqrt{\Sigma_\lambda}\,\epsilon$ with $\epsilon=0$. We keep $\beta>0$ as in the main experiments and use the same $\Sigma_\lambda$ (and hence, $\sigma_\lambda$) for a controlled comparison. We train this noise-free model with the objective in Eq.~(\ref{eq:training-loss}).

From a theoretical perspective, diffusion induced by noise leads to a convolution of probability distributions, and the influence of small-scale components is reflected in the statistical properties of large-scale components through the effective theory \citep{Carosso20,Cotler+Rezchikov23}. In other words, noise statistically represents the eliminated small-scale degrees of freedom. Theoretical details are provided in \cref{subsec:carosso-renormalization-group}. Here, we experimentally investigate the effect of noise.

\begin{figure*}[hb]
    \vskip 0.2in
    \begin{center}
        \includegraphics[width=\textwidth]{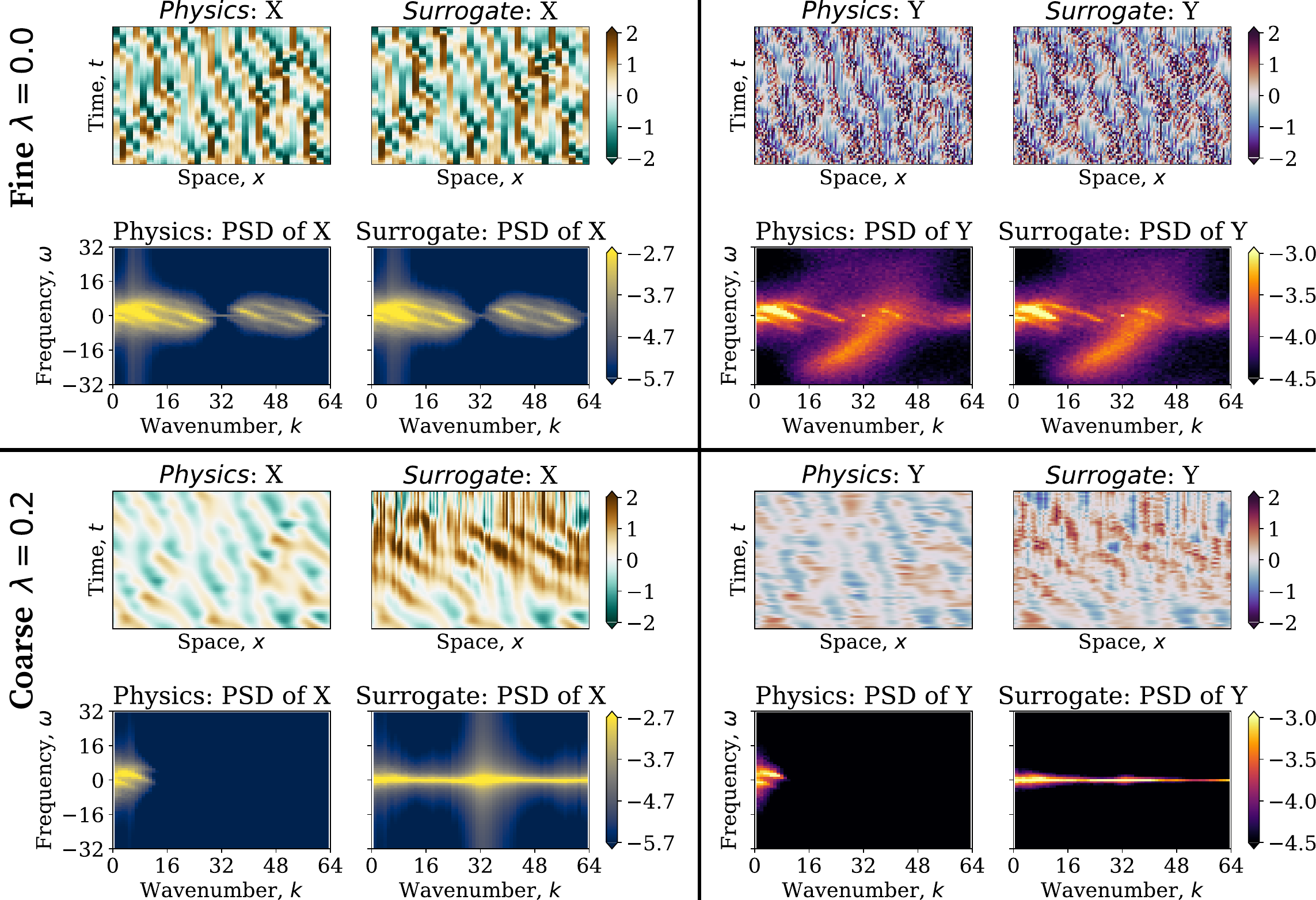}
        \caption{Simulation results for the Lorenz-96 model trained without noise injection ($\epsilon = 0$ in Eq.~(\ref{eq:ou-solution-spatiotemporal})). Top two rows: $\lambda = 0$; bottom two rows: $\lambda=0.2$. Left: slow variable $X$; right: fast variable $Y$. Each panel shows spatiotemporal evolution (top) and power spectral density (PSD; bottom). While prediction at $\lambda = 0$ remains stable, numerical instability increases at larger $\lambda$, demonstrating that noise is essential for stable coarse-grained simulation. PSDs are shown on a $\log_{10}$ scale. All quantities are non-dimensionalized.}
        \label{fig:appendix_lorenz96_simulation_ablation_without_noise}
    \end{center}
\end{figure*}

Even without noise, simulations are numerically stable when $\lambda \approx 0$; however, numerical instability increases and prediction fails as $\lambda$ becomes larger (Fig.~\ref{fig:appendix_lorenz96_simulation_ablation_without_noise}). It is known that injecting noise during training improves the numerical stability of surrogate models \citep{Pfaff+21,Sanchez+20}, and our results are consistent with this observation.

When no noise is injected during training, generation fails (not shown in detail). Specifically, the generated spatiotemporal samples look like noise, and the characteristic spatiotemporal patterns are not reproduced. Theoretically, noise induces a convolution of probability distributions, which corresponds to statistically integrating out the eliminated small-scale degrees of freedom \citep{Carosso20,Cotler+Rezchikov23}. Data generation via the reverse process can therefore be interpreted as reconstructing small-scale components from a statistically integrated representation \citep{Cotler+Rezchikov23}. When noise is not injected during training, such statistical integration does not occur, and the model fails to learn how to reconstruct small-scale components.

\begin{figure*}[tbh]
    \vskip 0.2in
    \begin{center}
        \includegraphics[width=\textwidth]{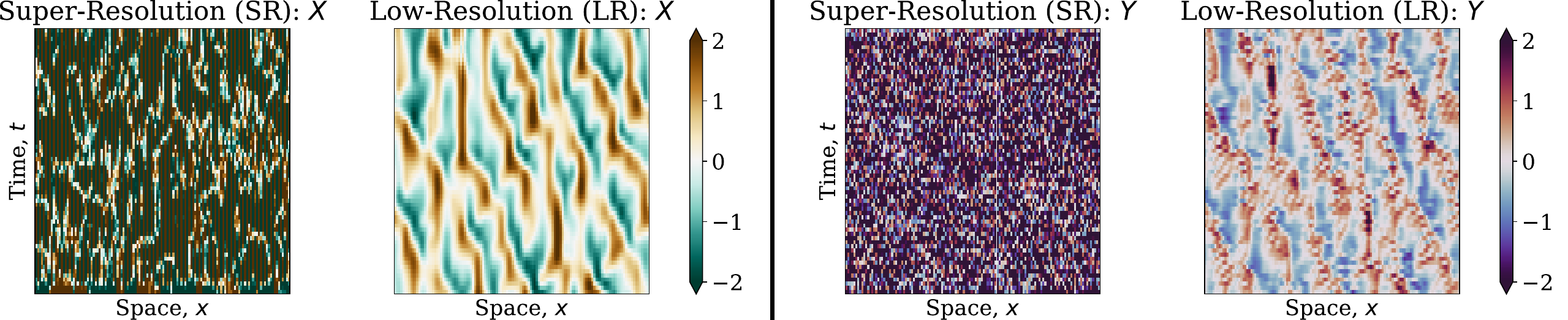}
        \caption{Super-resolution results for the Lorenz-96 model, without noise during training. The coarse-grained (low-resolution) result is obtained from the simulation at $\lambda=0.025$. Using this as input, super-resolution is performed following Algorithm~\ref{alg:generation}. All quantities are non-dimensionalized.}
        \label{fig:appendix_lorenz96_super_resolution_no_noise}
    \end{center}
    \vskip 0.4in
\end{figure*}

We further examine whether super-resolution is possible in the absence of noise. Choosing a relatively small diffusion scale for which coarse-grained (low-resolution) simulation is successful, $\lambda = 0.025$, we apply the super-resolution procedure. As expected, super-resolution also fails in this case (Fig.~\ref{fig:appendix_lorenz96_super_resolution_no_noise}). This result is consistent with the failure of unconditional spatiotemporal data generation and suggests that small-scale components cannot be reconstructed without statistical integration induced by noise.

\newpage

\subsection{Experimental Results with FNO}
\label{subsec:experimental-results-with-fno}

We evaluate our proposed method using the FNO architecture (\cref{subsec:neural-network-architecture}). For simulation at each $\lambda$ and unconditional generation, the FNO-based model yields results comparable to those obtained with the U-Net. Here, we focus on super-resolution, which combines forward-$t$ simulation and reverse-$\lambda$ sampling.

\begin{figure*}[t]
    \begin{center}
        \includegraphics[width=\textwidth]{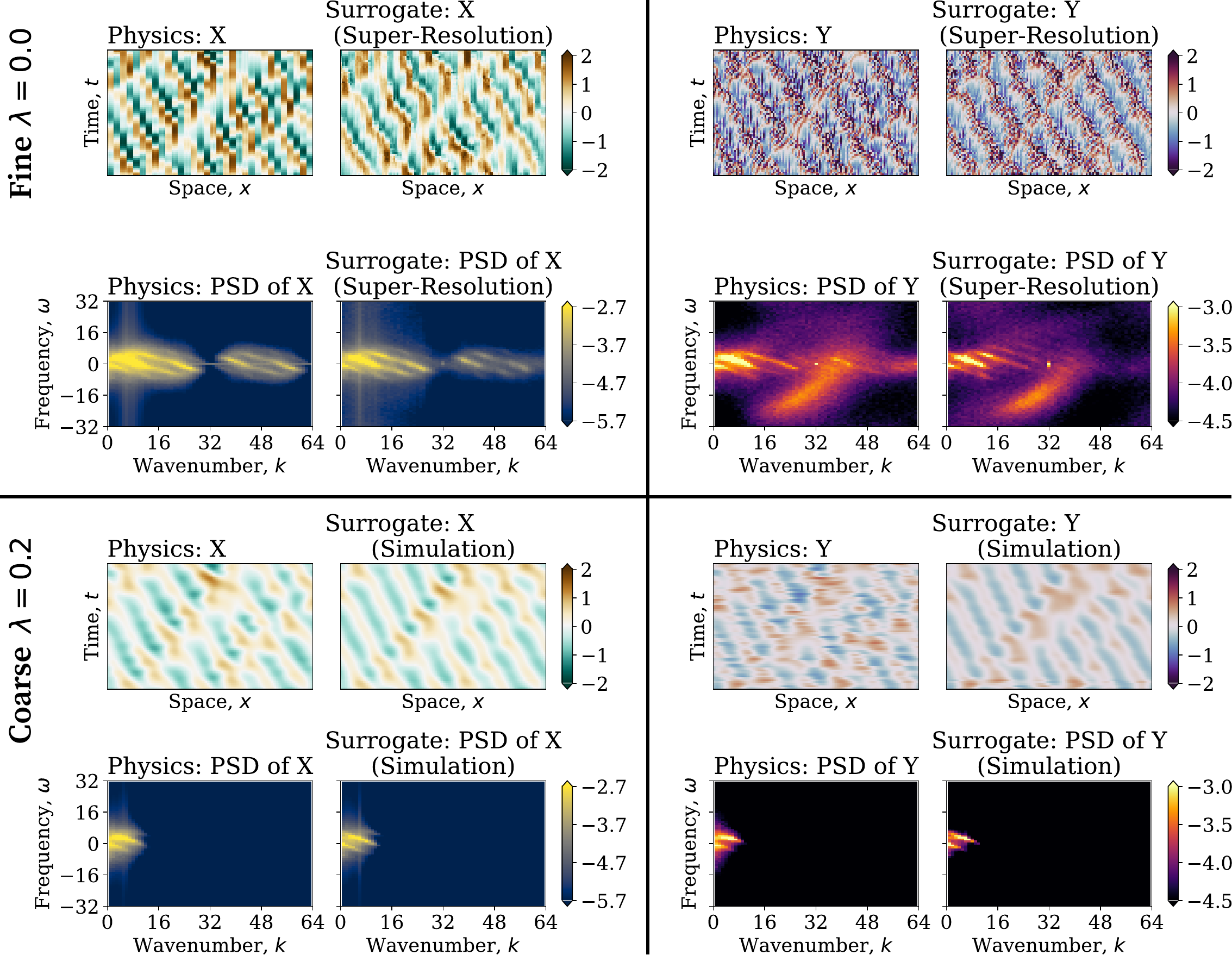}
        \caption{Super-resolution and simulation results for the Lorenz-96 model using an FNO-based surrogate. Top two rows: $\lambda = 0$; bottom two rows: $\lambda=0.2$. Left: slow variable $X$; right: fast variable $Y$. Each panel shows spatiotemporal evolution (top) and power spectral density (PSD; bottom), comparing physics-based simulation (left column) with the FNO surrogate (right column). The coarse-grained simulation ($\lambda=0.2$) captures large-scale dynamics, and the surrogate super-resolves this to $\lambda = 0$ (Algorithm~\ref{alg:generation}), recovering small-scale structure consistent with the physics-based reference. PSDs are shown on a $\log_{10}$ scale.}
        \label{fig:appendix_lorenz96_fno_simulation_and_super_resolution}
    \end{center}
\end{figure*}

Following the main text (\cref{subsec:experimental-results}), we perform simulation at $\lambda=0.2$ and apply reverse-$\lambda$ sampling to obtain super-resolved fields at $\lambda = 0$ (Fig.~\ref{fig:appendix_lorenz96_fno_simulation_and_super_resolution}). The coarse-grained (low-resolution) simulation at $\lambda=0.2$ reproduces the coarse-grained physical simulation well, capturing the fluctuations of large-scale components. The super-resolved result at $\lambda = 0$ recovers small-scale components and exhibits spatiotemporal patterns consistent with the physical simulation from the same initial condition. These results suggest that our proposed method is effective across different network architectures.

\end{document}